\documentclass[twocolumn]{aastex62}
\usepackage[]{threeparttable}
\usepackage{float}
\usepackage[caption = false]{subfig}
\usepackage{natbib}
\usepackage{appendix}
\usepackage{multirow}
\makeatletter
\newcommand\notsotiny{\@setfontsize\notsotiny{6}{7}}
\makeatother

\begin{document}
\title{Signatures of Tidal Disruption in Ultra-Faint Dwarf Galaxies:  A Combined \textit{HST}, \textit{Gaia}, and MMT/Hectochelle Study of Leo V  }

\correspondingauthor{Bur\c{c}in Mutlu-Pakdil}
\email{bmutlupakdil@as.arizona.edu}

\author[0000-0001-9649-4815]{BUR\c{C}\.{I}N MUTLU-PAKD\.{I}L}
\affil{Department of Astronomy/Steward Observatory, 933 North Cherry Avenue, 
Rm. N204, Tucson, AZ 85721-0065, USA}

\author[0000-0003-4102-380X]{DAVID J. SAND}
\affil{Department of Astronomy/Steward Observatory, 933 North Cherry Avenue, 
Rm. N204, Tucson, AZ 85721-0065, USA}

\author{MATTHEW G. WALKER}
\affil{Center for Cosmology, Department of Physics, Carnegie Mellon University, 5000 Forbes 
Avenue,Pittsburgh, PA 1521, USA}

\author{NELSON CALDWELL}
\affil{Center for Astrophysics, Harvard \& Smithsonian, 60 Garden Street, Cambridge, MA 02138, USA}

\author[0000-0002-3936-9628]{JEFFREY L. CARLIN}
\affil{LSST, 950 North Cherry Avenue, Tucson, AZ 85719, USA}

\author{MICHELLE L. COLLINS}
\affil{Department of Physics, University of Surrey, Guildford GU2 7XH, UK}

\author{DENIJA CRNOJEVI\'{C}}
\affil{University of Tampa, 401 West Kennedy Boulevard, Tampa, FL 33606, USA}

\author{MARIO MATEO}
\affil{Department of Astronomy, University of Michigan, Ann Arbor, MI 48109, USA}

\author{EDWARD W. OLSZEWSKI}
\affil{Department of Astronomy/Steward Observatory, 933 North Cherry Avenue, 
Rm. N204, Tucson, AZ 85721-0065, USA}

\author{ANIL C. SETH}
\affil{University of Utah, 115 South 1400 East Salt Lake City, UT 84112-0830, USA}

\author{JAY STRADER}
\affil{Department of Physics and Astronomy, Michigan State University,East Lansing, MI 48824, USA}

\author{BETH WILLMAN}
\affil{LSST and Steward Observatory, 933 North Cherry Avenue, Tucson, AZ 85721, USA}

\author{DENNIS ZARITSKY}
\affil{Department of Astronomy/Steward Observatory, 933 North Cherry Avenue, Rm. N204, 
Tucson, AZ 85721-0065, USA}
 
\begin{abstract}

The ultra-faint dwarf galaxy Leo~V has shown both photometric overdensities and kinematic members at large radii, along with a  tentative kinematic gradient, suggesting that it may have undergone a close encounter with the Milky Way. We investigate these signs of disruption through a combination of i) high-precision photometry obtained with the \textit{Hubble Space Telescope} (\textit{HST}), ii) two epochs of stellar spectra obtained with the Hectochelle Spectrograph on the MMT, and iii) measurements from  the \textit{Gaia} mission. Using the \textit{HST} data, we examine one of the reported stream-like overdensities at large radii, and conclude that it is not a true stellar stream, but instead a clump of foreground stars and background galaxies. Our spectroscopic analysis shows that one known member star is likely a binary, and challenges the membership status of  three others, including two distant candidates that had formerly provided evidence for overall stellar mass loss. We also find evidence that the proposed kinematic gradient across Leo~V might be due to small number statistics. We update the systemic proper motion of Leo~V, finding $(\mu_\alpha \cos\delta, \mu_\delta)= (0.009\pm0.560$,  $-0.777\pm0.314)$ mas yr$^{-1}$, which is consistent with its reported orbit that did not put Leo~V at risk of being disturbed by the Milky Way. These findings remove most of the observational clues that suggested Leo~V was disrupting, however, we also find new plausible member stars, two of which are located $>5$ half-light radii from the main body.  These stars require further investigation. Therefore, the nature of Leo~V still remains an open question.

\end{abstract}

\section{Introduction}\label{sec:intro}

Ultra-faint dwarf (UFD) galaxies are the oldest, smallest, most dark-matter dominated, and least chemically evolved stellar systems known. The study of these system has significant implications from the faint-end of the galaxy luminosity function \citep{Koposov2009} to the validity of cosmological models \citep[e.g.,][]{BullockBoylan2017,Pawlowski2017,Kim2017,TulinYu2018}, including the nature of dark matter \citep[e.g.,][]{GeringerSameth2015,Calabrese2016,Jethwa2018,Errani2018,Bozek2019,Robles2019a,Robles2019} and the formation of the first galaxies \citep[e.g.,][]{Bovill2011,Wheeler2015}.

However, there are several observational challenges in understanding such faint systems. Given the presence of only a handful of bright stars, it has been very difficult to study them spectroscopically. Even under the assumption of minimal contamination by binary and foreground stars, they are largely doomed to suffer from small number statistics. The presence or absence of dark matter, and thus whether they are indeed galaxies, remains unclear in some of the recently discovered systems (e.g., Triangulum~II: \citealt{Kirby2015,Martin2016,Kirby2017,Carlin2017,Ji2019}, Tucana~III: \citealt{Simon2017,Li2018}, Tucana~V: \citealt{Conn2018}, Sagittarius~II: \citealt{MutluPakdil2018,Longeard2019}). The existing dynamical analyses rely heavily on the assumptions of dynamical equilibrium.
Yet, it has frequently been suggested that several UFDs have been affected by Galactic tides \citep[e.g.,][]{Belokurov2006,Zucker2006,NiedersteOstholt2009,Belokurov2009,Sand2009,Munoz2010,Sand2012,Kirby2013,Roderick2015,Simon2017,Collins2017,Garling2018}. Thanks to \textit{Gaia}, the orbits of the UFDs are now constrained \citep[e.g.,][]{Simon2018,Fritz2018}, and Tucana~III is unambiguously suffering substantial stripping \citep{Shipp2018,Li2018,Erkal2018}, with its orbital pericenter of only $\sim3$~kpc \citep{Simon2018,Fritz2018}. Several UFDs are found to have orbits that might bring them close to the inner regions of the Milky Way, making them likely to be tidally disturbed: Bo\"{o}tes~III \citep{Carlin2018}, Willman~1, Segue~1, Triangulum~II \citep{Fritz2018}, Crater~II, Hercules \citep{Fritz2018,Fu2019}, and Draco~II \citep{Fritz2018,Longeard2018}. On the other hand, Leo~V shows signs of tidal disturbance but does not have an orbit that seem to put it at risk of being tidally disturbed by the Milky Way.

In this paper, we focus on Leo~V \citep[discovered in][]{Belokurov2008} and critically assess previously observed signatures of tidal influence in combination with new observations.  Ground-based observations suggest Leo~V has an extended morphology, with a highly elongated shape (ellipticity $\sim0.5$) and stellar overdensities outside of several half-light radii \citep{Sand2012}. \citet{Walker2009} found that the likely blue horizontal branch member distribution was more extended than the bulk of red giant branch stars. Leo~V also displays a tentative velocity gradient, which may indicate it is in an advanced stage of dissolution \citep{Collins2017}. These ubiquitous hints for tidal effects among distant dwarfs like Leo~V is particularly surprising because they can only experience tidal stripping if their orbits are extremely eccentric, bringing them within 10$-$20~kpc of the Galactic center \citep{Simon2019}. Based on \textit{Gaia} proper motions of five central Leo~V member candidates reported in \citet{Walker2009}, \citet{Fritz2018} found it unlikely for Leo~V to have such an orbit. The authors estimated an orbital pericenter of $165^{+14}_{-126}$ ($168^{+12}_{-104}$)~kpc for Leo~V, assuming a Milky Way dark matter halo with virial mass $1.6\times10^{12} M_{\odot}$ ($0.8\times10^{12} M_{\odot}$). Therefore, it is crucial to verify whether the photometric tidal features observed around Leo~V are true tidal material$-$stripped stars$-$or whether clustered background galaxies are masquerading as stellar debris and, by extension, whether the kinematic gradients are real or due to small number statistics. The confirmation of even a small number of tidally stripped stars would imply much more global mass loss, in both dark matter and stars.

Here we investigate the signs of tidal disruption in Leo~V through a combination of high-precision photometry obtained with the \textit{HST}/ACS and WFC3, two epochs of stellar spectra obtained with the MMT/Hectochelle Spectrograph, and proper motion measurements from \textit{Gaia}~DR2. We describe our observations and data reduction in Section~\ref{sec:obs}. We revisit the distance to Leo~V and present its color-magnitude diagram (CMD) in comparison to those of Leo~IV and M92 in Section~\ref{sec:cmddist}. Using \textit{HST} data, we investigate the true nature of the tentative candidate debris stream of Leo~V in Section~\ref{sec:nugget}. In Section~\ref{sec:members}, we present out new spectroscopic results, along with the relevant \textit{Gaia}~DR2 data, to search for any signs of tidal disturbance, while updating the proper motion measurement of Leo~V. Finally, we summarize our key results in Section~\ref{sec:conclusion}.

\section{Observations and Data Reduction}\label{sec:obs}

\subsection{\textit{HST} Imaging}

\begin{figure}
\centering
\includegraphics[width=\columnwidth]{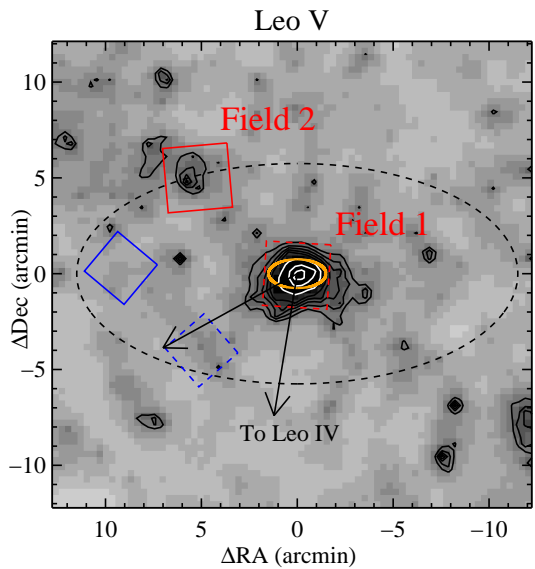}
\caption{Our observing strategy: the smoothed matched-filter map of Leo~V from the ground-based imaging of \citet{Sand2012} with half-light radii marked in orange (for Leo~V, this value is comparable to the smoothing size). Dashed boxes, labeled as Field~1, represent the archived \textit{HST}/ACS and WFC3 imaging which we utilized for our analysis. The solid boxes, labeled as Field~2, are our \textit{HST}/ACS (red) pointing along the putative streams (far from the main body of Leo~V) and WFC3 (blue) parallel. The black dashed ellipse is the approximate isodensity contour at the radius of our data ($r\sim10\times$ half-light radii), and the arrow shows the direction towards the Galactic Center. An additional arrow is shown to highlight the direction to Leo~IV. \label{fig:obsplan}}
\end{figure}

We have obtained deep optical observations of the candidate debris stream of Leo~V, which was identified in ground-based (Magellan/Megacam) data \citep{Sand2012}, using the F606W and F814W filters on the ACS (HST-GO-15182; PI: D. Sand). Table~\ref{tab:obslog} presents the log of the observations. A standard 4-point dither pattern was used to achieve 0.5 pixel sampling. The image depth was chosen to be consistent with the already-archived central pointing of Leo~V from HST-GO-14770 (PI: S. Sohn). Coordinated parallel observations with WFC3 were used to provide control CMDs to assess the impact of foreground star/background galaxy contamination. Our observational strategy is outlined in Figure~\ref{fig:obsplan}, in which we refer to the central pointing of Leo~V as Field~1 and the candidate debris stream as Field~2. We made use of archived Leo~IV data as well (HST-GO-12549; PI: T. Brown) for comparison purposes. Leo~IV is very close to Leo~V in both location (position and distance) and radial velocity, implying a possible common origin for both galaxies \citep[][but see \citealt{Sand2010,Jin12}]{Belokurov2008,deJong2010,Blana2012,Munoz2018}. 

\begin{table}[htp!]
\centering
\notsotiny
\caption{Observation Log and Field Completeness of Leo~V-related fields (Fields~1-2, see Figure~\ref{fig:obsplan}) and Leo~IV.} \label{tab:obslog}
\begin{tabular}{lccccc}
\tablewidth{0pt}
\hline
\hline
Field Name    & Camera &  Filter & Exp  & 50\%  & 90\%  \\
{}            &  {}    &  {}     & (s)  & (mag) & (mag) \\
\hline
Leo~V Field~1 & ACS   &  F606W & 4557  & 27.55 & 26.48  \\
{(Main Body)} & ACS   &  F814W & 4565  & 27.29 & 26.67  \\
{}            & WFC3  &  F606W & 4596  & 27.24 & 26.29  \\
{}            & WFC3  &  F814W & 4605  & 27.07 & 26.66  \\
\hline
Leo~V Field~2 & ACS   &  F606W & 5174  & 27.59 & 26.50  \\
{(Stream Candidate)}  & ACS   &  F814W & 5174  & 27.30 & 26.67    \\
{}            & WFC3  &  F606W & 5264  & 27.34 & 26.32  \\
{}            & WFC3  &  F814W & 5264  & 27.14 & 26.58  \\
\hline
Leo~IV        & ACS   &  F606W & 20540 & 28.35 & 27.29  \\
{}            & ACS   &  F814W & 20540 & 28.13 & 27.52  \\
\hline
\end{tabular}
\end{table}

\begin{table*}[tbp]
\caption{Photometry of Field~1-ACS: Central Pointing of Leo~V.} \label{tab:leov}
\begin{minipage}[b]{0.95\linewidth}\centering
\begin{tabular}{ccccccccc}
\tablewidth{0pt}
\hline
\hline
Star No. & $\alpha$      &  $\delta$      & F606W     & $\delta$(F606W) & $A_{F606W}$ & F814W & $\delta$(F814W)  & $A_{F814W}$ \\
{}       & (deg J2000.0) &  (deg J2000.0) & (mag) & (mag)     & (mag)   & (mag) & (mag)     & (mag)   \\
\hline
0 & 172.80600 & 2.2229825 & 20.00 & 0.01 & 0.07 & 17.62 & 0.01 & 0.04\\
1 & 172.76334 & 2.2021950 & 19.32 & 0.01 & 0.07 & 17.67 & 0.01 & 0.04\\
2 & 172.81251 & 2.2487917 & 19.10 & 0.01 & 0.07 & 18.25 & 0.01 & 0.04\\
3 & 172.76198 & 2.2207080 & 19.59 & 0.01 & 0.07 & 18.78 & 0.01 & 0.04\\
4 & 172.79412 & 2.2359477 & 19.91 & 0.01 & 0.07 & 19.04 & 0.01 & 0.04\\
\hline
\end{tabular}
   \begin{tablenotes}
      \small
      \item (This table is available in its entirety in a machine-readable form in the online journal. A portion is shown here for guidance regarding its form and content.)  
    \end{tablenotes}
\caption{Photometry of Field~2-ACS: Candidate Debris Stream of Leo~V.} \label{tab:nugget}
\begin{tabular}{ccccccccc}
\tablewidth{0pt}
\hline
\hline
Star No. & $\alpha$      &  $\delta$      & F606W     & $\delta$(F606W) & $A_{F606W}$ & F814W & $\delta$(F814W)  & $A_{F814W}$ \\
{}       & (deg J2000.0) &  (deg J2000.0) & (mag) & (mag)     & (mag)   & (mag) & (mag)     & (mag)   \\
\hline
0 & 172.84364 & 2.2901851 & 20.57 & 0.01 & 0.07 & 18.61 & 0.01 & 0.04\\
1 & 172.84970 & 2.3302994 & 20.91 & 0.01 & 0.07 & 18.72 & 0.01 & 0.04\\
2 & 172.86505 & 2.3338706 & 21.56 & 0.01 & 0.07 & 19.15 & 0.01 & 0.04\\
3 & 172.88719 & 2.3170981 & 20.16 & 0.01 & 0.07 & 19.52 & 0.01 & 0.04\\
4 & 172.84669 & 2.3246526 & 20.34 & 0.01 & 0.07 & 19.57 & 0.01 & 0.04\\
\hline
\end{tabular}
   \begin{tablenotes}
      \small
      \item (This table is available in its entirety in a machine-readable form in the online journal. A portion is shown here for guidance regarding its form and content.)
    \end{tablenotes}
\caption{Photometry of Field~1-WFC3: Parallel Observations of Leo~V with WFC3.} \label{tab:leovwfc3}
\begin{tabular}{ccccccccc}
\tablewidth{0pt}
\hline
\hline
Star No. & $\alpha$      &  $\delta$      & F606W     & $\delta$(F606W) & $A_{F606W}$ & F814W & $\delta$(F814W)  & $A_{F814W}$ \\
{}       & (deg J2000.0) &  (deg J2000.0) & (mag) & (mag)     & (mag)   & (mag) & (mag)     & (mag)   \\
\hline
0 & 172.86426 & 2.1426387 & 21.02& 0.01& 0.07& 20.50 & 0.01& 0.04\\
1 & 172.85726 & 2.1384186 & 21.39& 0.01& 0.07& 20.78 & 0.01& 0.04\\
2 & 172.85070 & 2.1411125 & 21.43& 0.01& 0.07& 20.81 & 0.01& 0.04\\
3 & 172.86579 & 2.1721304 & 22.15& 0.01& 0.07& 20.82 & 0.01& 0.04\\
4 & 172.84163 & 2.1625177 & 22.06& 0.01& 0.07& 21.33 & 0.01& 0.04\\
\hline
\end{tabular}
   \begin{tablenotes}
      \small
      \item (This table is available in its entirety in a machine-readable form in the online journal. A portion is shown here for guidance regarding its form and content.)
    \end{tablenotes}
\caption{Photometry of Field~2-WFC3: Parallel Observations of Candidate Debris Stream of Leo~V with WFC3.} \label{tab:nuggetwfc3}
\begin{tabular}{ccccccccc}
\tablewidth{0pt}
\hline
\hline
Star No. & $\alpha$      &  $\delta$      & F606W     & $\delta$(F606W) & $A_{F606W}$ & F814W & $\delta$(F814W)  & $A_{F814W}$ \\
{}       & (deg J2000.0) &  (deg J2000.0) & (mag)     & (mag)           & (mag)       & (mag) & (mag)            & (mag)   \\
\hline
0 & 172.93332 & 2.2278148 & 21.27& 0.01& 0.07& 19.44& 0.01& 0.04\\
1 & 172.92149 & 2.2258288 & 22.56& 0.01& 0.07& 20.26& 0.01& 0.04\\
2 & 172.95559 & 2.2282597 & 22.98& 0.01& 0.07& 21.33& 0.01& 0.04\\ 
3 & 172.92321 & 2.2362039 & 22.63& 0.01& 0.07& 21.41& 0.01& 0.04\\
4 & 172.94284 & 2.2310564 & 22.45& 0.01& 0.07& 21.55& 0.01& 0.04\\
\hline
\end{tabular}
    \begin{tablenotes}
      \small
      \item (This table is available in its entirety in a machine-readable form in the online journal. A portion is shown here for guidance regarding its form and content.)
    \end{tablenotes}
\end{minipage}    
\end{table*}

We performed point-spread function photometry on the pipeline-produced flat-fielded (FLT) images using the latest version (2.0) of DOLPHOT \citep{Dolphin2002}, an updated version of HSTPHOT \citep{Dolphin2000}, largely using the recommended prescriptions on each camera. Drizzled (DRZ) images were used only as an astrometric reference frame; all photometry was performed on the FLT images. The catalogs were cleaned of background galaxies and stars with poor photometry, rejecting outliers in (sharpness$_{F606W}+$sharpness$_{F814W})^2 < 0.1$, (crowd$_{F606W}+$crowd$_{F814W}) < 0.08$, signal-to-noise ratio $> 5$, roundness $< 1.5$, and object-type $\leq 2$ in each filter. We corrected for Milky Way extinction on a star by star basis using the \citet{Schlegel1998} reddening maps with the coefficients from \citet{Schlafly2011}. Tables~\ref{tab:leov}-\ref{tab:nuggetwfc3} present our final catalogs, which include magnitudes (uncorrected for extinction) along with their DOLPHOT uncertainty, as well as the Galactic extinction values derived for each star. The extinction-corrected photometry is used throughout this work, and the CMDs for the main bodies of both Leo~IV and Leo~V are displayed in Figure~\ref{fig:compare}.

We derived completeness and photometric uncertainties using $\sim$50,000 artificial star tests per pointing, with the same photometric routines used to create the photometric catalogs. While the 50\% completeness limits of Fields~1-2 are 27.3~mag in F814W for the ACS fields, the Leo~IV limit reaches 28.1~mag (see Table~\ref{tab:obslog}), and the photometric uncertainties are accordingly smaller for the latter (see Figure~\ref{fig:compare}).  

\subsection{Spectroscopy}\label{subsec:spectra}

Here we present two epochs of Leo~V spectroscopy obtained with the Hectochelle multi-object fiber spectrograph \citep{Hecto} at the MMT telescope. Hectochelle has a 1~deg diameter field of view and can achieve $\sim$1~km~s$^{-1}$ velocity precision for high signal-to-noise data \citep{Walker2015}; the effective resolution for our setup was $R$ $\approx$32000. Both data sets were taken with the ``RV31'' filter in place, giving a spectral range of 5150--5300 \AA, which contains the prominent Mg b triplet. The first epoch of observations was taken on 2008 May 28 and 30, and was reported in \citet{Walker2009}, while the second epoch of observations was taken on 2009 March 03 (UT times are used throughout this work). Candidate Leo~V stars were selected based both on their proximity to the center of Leo~V and on their position along the CMD; plausible red giant branch stars were preferentially selected. A preference for some repeat observations of clear Leo~V stars in the 2008 dataset was made when observing the 2009 dataset in order to assess the impact of binary stars. We refer the reader to previous work with Hectochelle and the study of Milky Way dwarf galaxies for further details on the observational setup and data reduction \citep[e.g.][]{Mateo08,Belokurov2009,Caldwell17}.

We revisit the first epoch of previously published 2008 data alongside the 2009 dataset in order to analyze both in a uniform, robust fashion; although some detailed results have changed, the current results are largely consistent with that presented in \citet{Walker2009}. The sky subtracted spectra were analyzed following the procedure of \citet{Walker2015}, using a set of synthetic stellar templates \citep{Lee08a,Lee08b} to obtain Bayesian inferences on the line-of-sight velocity ($v$), effective temperature ($T_{eff}$), surface gravity (log $g$) and metallicity ([Fe/H]). We also enforce the quality-control criteria recommended by \citet{Walker2015}, discarding observations where the probability distribution function for $v$ is non-Gaussian. To do this, we retain only objects with $v$ uncertainty $\delta v<$ 5~km~s$^{-1}$, a skewness $|S|$$\leq$1 and kurtosis $|K|$$\leq$1. Finally, we apply zero-point offsets to our mean and variance measurements based on direct measurements of twilight sky spectra in comparison to solar values. The offset values applied are identical to those obtained for the Draco observations reported by \citet{Walker2015}.

There are 140 stellar observations that pass our quality-control criteria; 92 from the 2008 data set and 48 from the 2009 data set.  In total, there are 14 stars that have repeat measurements between the two epochs. Table~\ref{tab:hectoleo5} lists the measured parameters ($v$, $T_{eff}$, log $g$, and [Fe/H]) for each.  

\begin{table*}[tb!]
\caption{Hectochelle Stellar Spectroscopy of Leo V}\label{tab:hectoleo5}
\begin{minipage}[b]{0.95\linewidth}\centering
\begin{tabular}{lccccccc}
\tablewidth{0pt}
\hline
\hline
ID & R.A. & Dec & HJD & $v$ & $T_{eff}$ & log[$g$/(cm s$^{-2}$)] & [Fe/H] \\
 {} & (deg) & (deg) & -2450000.0 (days) & (km s$^{-1}$) & (K) & (dex) & (dex)\\
\hline
LeoV-0 & 172.52875 &  2.12367 &  4614.7 & -4.3$\pm$2.6 & 4962$\pm$523 & 3.81$\pm$0.71 & -2.57$\pm$0.61 \\
LeoV-1 & 172.76200 &  2.22075 &  4614.7 & 174.3$\pm$1.7 & 4784$\pm$418 & 1.70$\pm$0.78 & -2.76$\pm$0.49 \\
LeoV-2 & 172.45170 &  2.10953 &  4614.7 & -58.4$\pm$1.6 & 5616$\pm$778 & 2.18$\pm$1.00 & -1.29$\pm$0.72 \\
LeoV-3 & 172.56208 &  2.24169 &  4614.7 & 162.4$\pm$1.9 & 4476$\pm$264 & 3.50$\pm$0.69 & 0.55$\pm$0.32 \\
LeoV-4 & 172.71446 &  2.23375 &  4614.7 & 70.9$\pm$0.9 & 4405$\pm$112 & 4.58$\pm$0.29 & -0.51$\pm$0.16 \\
\hline
\end{tabular}
    \begin{tablenotes}
      \small
      \item (This table is available in its entirety in a machine-readable form in the online journal. A portion is shown here for guidance regarding its form and content.)
    \end{tablenotes}
\end{minipage}    
\end{table*}

\section{Distance and Color-Magnitude Diagram} \label{sec:cmddist}

\begin{figure*}[hbt!]
\centering
\includegraphics[width=2\columnwidth]{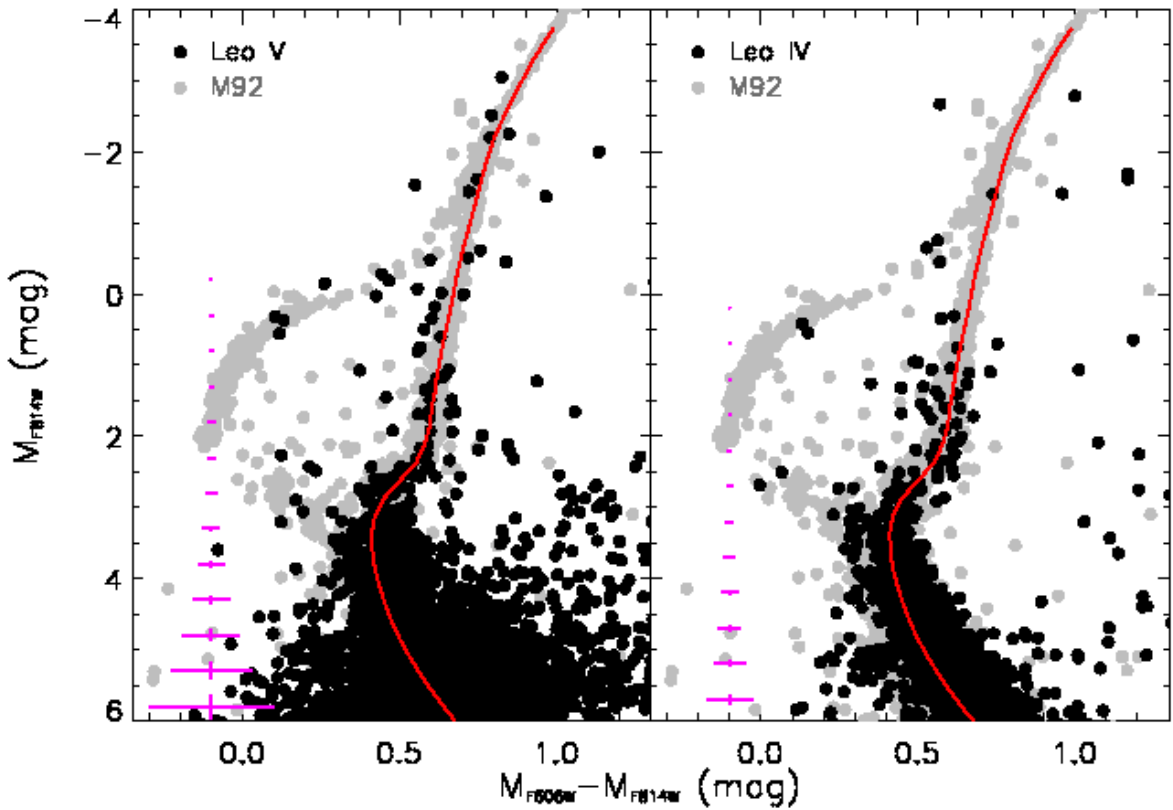}
\caption{A comparison  of the CMDs of Leo~V (left) and Leo~IV (right), relative to M92 (gray points). The figure only includes stars from the central ACS pointing of each object. Overplotted as a red line is our M92 fiducial sequence. The CMDs show a close agreement between Leo~IV, Leo~V and M92. 
\label{fig:compare}}
\end{figure*}

To fully understand the properties of Leo~V, it is important to make a comparison with similar stellar systems. Given the relative distance and velocity, Leo~IV is an ideal target for this. Also, the Galactic globular cluster M92 has been commonly used as a reference population for the UFD galaxies because it is one of the most ancient, metal-poor, and well-studied star clusters known. The reported mean metallicity from medium-resolution spectra is $-2.48\pm0.21$ for Leo~V \citep{Collins2017} and $-2.54\pm0.86$ for Leo~IV \citep{Kirby2011}. Numerous high resolution studies have derived metallicities ranging from $-2.4<$[Fe/H]$<-2.1$ for M92 \citep[e.g.,][]{Sneden2000,Behr2003,Carretta2009}, though others have presented evidence for [Fe/H]$<-2.5$ in individual stars of M92 \citep[e.g.,][]{Peterson1990,King1998,Roederer2011}. Therefore, M92 provides an important empirical fiducial for the stellar populations of both Leo~V and Leo~IV. We revisit the distance measurements for Leo~V and Leo~IV by making a comparison to the ridgeline of M92. The details of the M92 \textit{HST} photometry and our derivation of its fiducial sequence are described in  Appendix~\ref{sec:AppendixA}. Note that we implement the extinction corrections for both Leo~IV and M92 using the same method described in Section~\ref{sec:obs}, with an average $E(B-V)$ of $0.025$~mag and $0.022$~mag, respectively. 

We determine the distance modulus of Leo~V by counting the number of stars consistent with the M92 fiducial, as described in \citet[][see also \citealt{Walsh2008}]{MutluPakdil2018}. We assume a distance modulus of $m-M=14.62$~mag for M92 as in \citet{Brown2014}, taking the mean of measurements from \citet[$14.60\pm0.09$ mag]{Paust2007}, \citet[$14.62\pm0.1$ mag]{DelPrincipe2005}, and \citet[$14.65\pm0.1$ mag]{Sollima2006}. The fiducial is shifted through 0.025 mag intervals in ($m-M$) from 21.0 to 22.0 mag in F814W, a plausible range of distance moduli for Leo~V. In each step, we count the number of stars consistent with the fiducial, taking into account photometric uncertainties. We also account for background stars by running the identical procedure in parallel over Field~2. Although our Field~2 pointing targets a stream candidate of Leo~V, it is placed far from its main body, hence we expect to find fewer dwarf stars in that field. We count the number of Field~2 stars consistent with the fiducial and then subtract this number from that derived in Field~1. We derive the best-fit distance modulus to be where the fiducial gives the maximum number of dwarf stars. We use a 100 iteration bootstrap analysis to determine the uncertainties, and find $21.25\pm0.08$ mag (D$=178\pm7$~kpc, see Table~\ref{tab:leovstr}). 

We also derive a distance modulus using the possible blue horizontal-branch stars (HBs) of Leo~V within our field of view (four stars). We fit to the HB sequence of M92 by minimizing the sum of the squares of the difference between the data and the fiducial. The best-fit distance modulus from HBs is $21.20\pm0.08$ mag, where the associated uncertainty comes from jackknife resampling, which is consistent with our initial finding. Due to the small number of HB stars within our field of view, we opt to adopt $21.25\pm0.08$ mag as our final distance modulus value for Leo~V. Our measurements also agree well with the distance estimation from RR Lyrae stars ($21.19\pm0.06$, \citealt{Medina2017}), and are consistent with the \citet{Sand2012} HB-derived distance ($21.46\pm0.16$) within uncertainties.

For comparison purposes, we made use of the \textit{HST} archival data of Leo~IV. \citet{Brown2014} determined the distance to Leo~IV ($m-M=21.12\pm0.07$) by using the same ACS data, but their approach in both photometry and distance measurement was different than ours. After performing photometry using DOLPHOT, we derive the Leo~IV distance using the same method described above, employing our new M92 fiducial. The resulting distance modulus is $m-M=20.85\pm0.07$ ($D$=148$\pm5$~kpc), which is consistent with the distance estimate from its RR Lyrae stars ($20.94\pm0.07$, \citealt{Moretti2009}). We also measure the distance modulus using two possible blue HB stars of Leo~IV within the field of view, and find $21.02\pm0.13$ mag, which is consistent with our adopted value.  

\begin{table}[htp!] 
\centering
\caption{Structural Properties of Leo~V} \label{tab:leovstr}
\begin{tabular}{lcc}
\tablewidth{0pt}
\hline
\hline
Parameter & Leo~V & Ref. \\
\hline
R.A. (deg)           & $172.78404$     & 1 \\
Dec. (deg)           & $2.22205$       & 1 \\
$M_{V}$ (mag)        & $-4.4\pm0.4$    & 1 \\     
$r_{h}$ (arcmin)     & $1.14\pm0.53$   & 1 \\     
$r_{h}$ (pc)   	     & $65\pm30$       & 1 \\  
Ellipticity          & $0.52\pm0.26$   & 1 \\ 
Position Angle (deg) & $90\pm10$       & 1 \\
$m-M$ (mag)          & $21.25\pm0.08$  & 2 \\
Distance (kpc)       & $178\pm7$       & 2 \\  
$\langle E(B-V)\rangle$ &  0.027       & 2 \\
$\mu_\alpha \cos\delta$ (mas yr$^{-1}$) & $0.009\pm0.560\pm0.057$ & 2\\ 
$\mu_\delta$ (mas yr$^{-1}$) & $-0.777\pm0.314\pm0.057$ & 2\\ 
\hline
\end{tabular}
  \begin{tablenotes}
      \small
      \item Notes: Last column is for references: (1) \citet{Sand2012} and (2) this work. 
    \end{tablenotes}
\end{table}

Figure~\ref{fig:compare} shows the final CMDs of Leo~V (Field~1-ACS) and Leo~IV, relative to M92. Magenta error bars are the mean photometric errors determined from artificial stars, and they are plotted at an arbitrary color for convenience. As one would expect from the image depths, the photometric errors are larger for Leo~V.  
As derived from the same ACS data, our CMD of Leo~IV is expected to be very similar to the one in \citet{Brown2014}. The authors compared their Leo~IV CMD with their M92 fiducial \citep{Brown2005}, along with five other ultra-faint dwarf galaxies (i.e., Bo\"{o}tes~I, Canes Venatici~II, Coma Berenices, Hercules, and Ursa Major~I). In addition to overall good agreement with M92, their CMDs show the presence of a stellar population in Leo~IV (and the other dwarfs) that is bluer and brighter than the M92 ridge line near the turnoff. As a plausible explanation, the authors suggested that these stars were more metal poor than those in M92. In Figure~\ref{fig:compare}, however, our CMD shows a close agreement between Leo~IV and M92 stars, including near the turnoff. It is worth mentioning that there is a relative color shift between our CMD and theirs due to the difference between the adopted reddening values. \citet{Brown2014} derived the distance and extinctions from fits to the ACS data and adopted $E(B-V)= 0.08$~mag for Leo~IV, which is much higher than our adopted value ($E(B-V)= 0.025$~mag on average), which comes from the \citet{Schlafly2011} extinction derived from the \citet{Schlegel1998} reddening maps. This explains the presence of a bluer and brighter star population in their CMD, which we do not think is real. 

Figure~\ref{fig:compare} clearly shows the globular cluster M92 is a nice fit to ultra-faint dwarf galaxies that are dominated by ancient metal-poor populations. Both Leo~V and Leo~IV display a very close agreement with M92. As their CMDs are all very similar to one another, Leo~IV, Leo~V and M92 should have similar stellar populations and star formation histories. On the other hand, based on deep ground imaging from \citet{Sand2012}, Leo~V shows signs of stream-like overdensities at large radii, while Leo~IV appears to show no signs of tidal debris \citep{Sand2010}. We further investigate the possible tidal disruption  of Leo~V in the following sections.

\section{Is the stream-like structure seen in ground-based observations real?}\label{sec:nugget}

\begin{figure*}[tbh!] 
\centering
    \includegraphics[width=1.25\columnwidth]{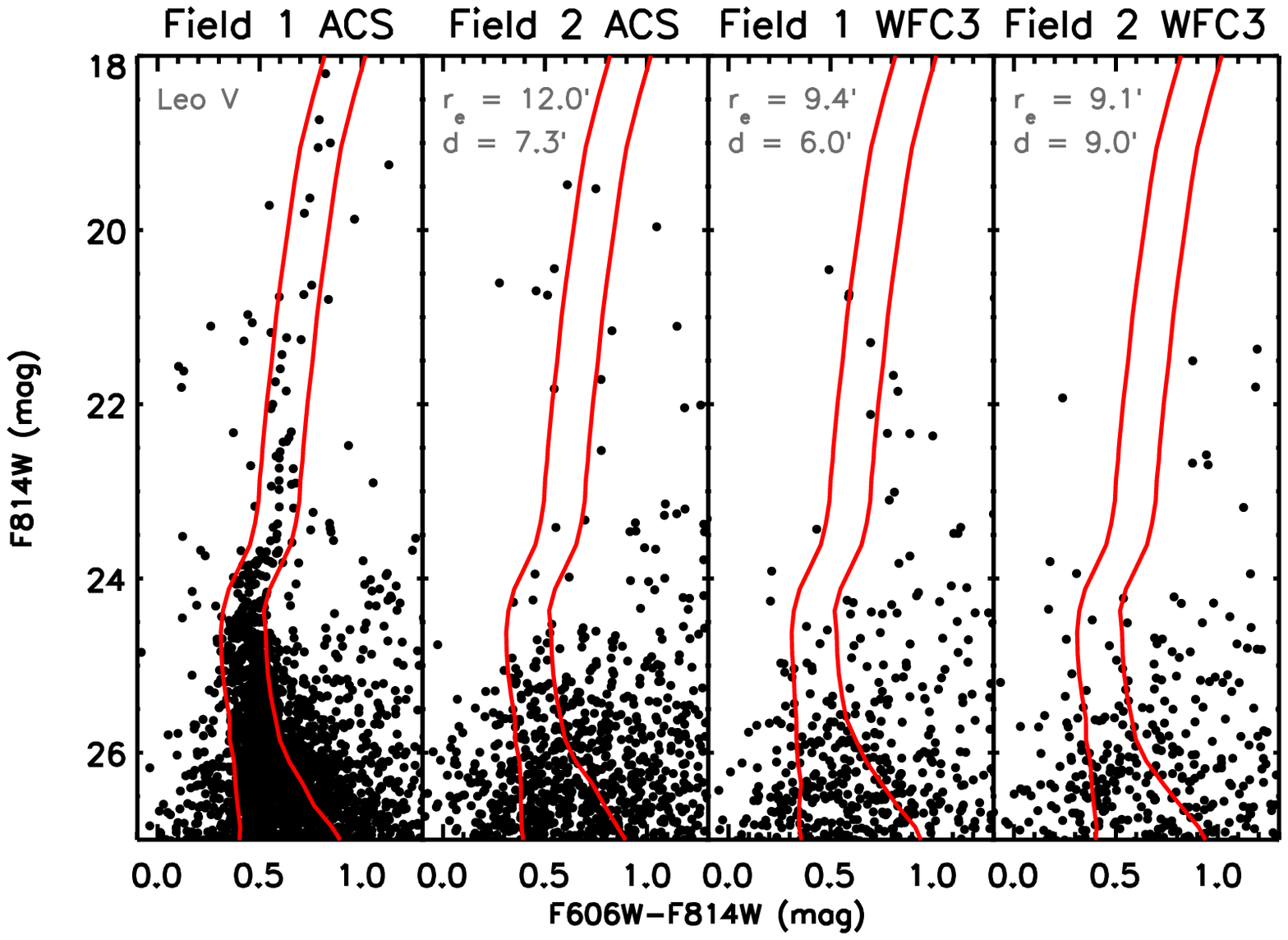}
    \caption{CMD of all the data within each \textit{HST} field. Red lines highlight the region selected by our M92 fiducial filter. \textbf{Far-left:} Field~1-ACS is the central pointing of Leo~V. \textbf{Center-left:} Field~2-ACS, centered on the candidate debris stream of Leo~V. \textbf{Center-right:} Field~1-WFC3 is the parallel observations of the central Leo~V field, taken with WFC3. \textbf{Far-right:} Field~2-WFC3 is the parallel observations of the candidate debris stream field of Leo~V, taken with WFC3. For the three offset fields, we list the distance, $d$,  from the center of Field~1 (Leo~V) and the elliptical radius, $r_{e}$, assuming an ellipticity of $\sim$0.5 for Leo~V \citep{Sand2012}. Note that the ellipticity of Leo~V is not well-constrained, and the Field~2-ACS pointing is at similar Leo~V-centric radii to the two WFC3 parallel fields within the uncertainties. \label{fig:nuggetinspect}}
    \includegraphics[width=1.25\columnwidth]{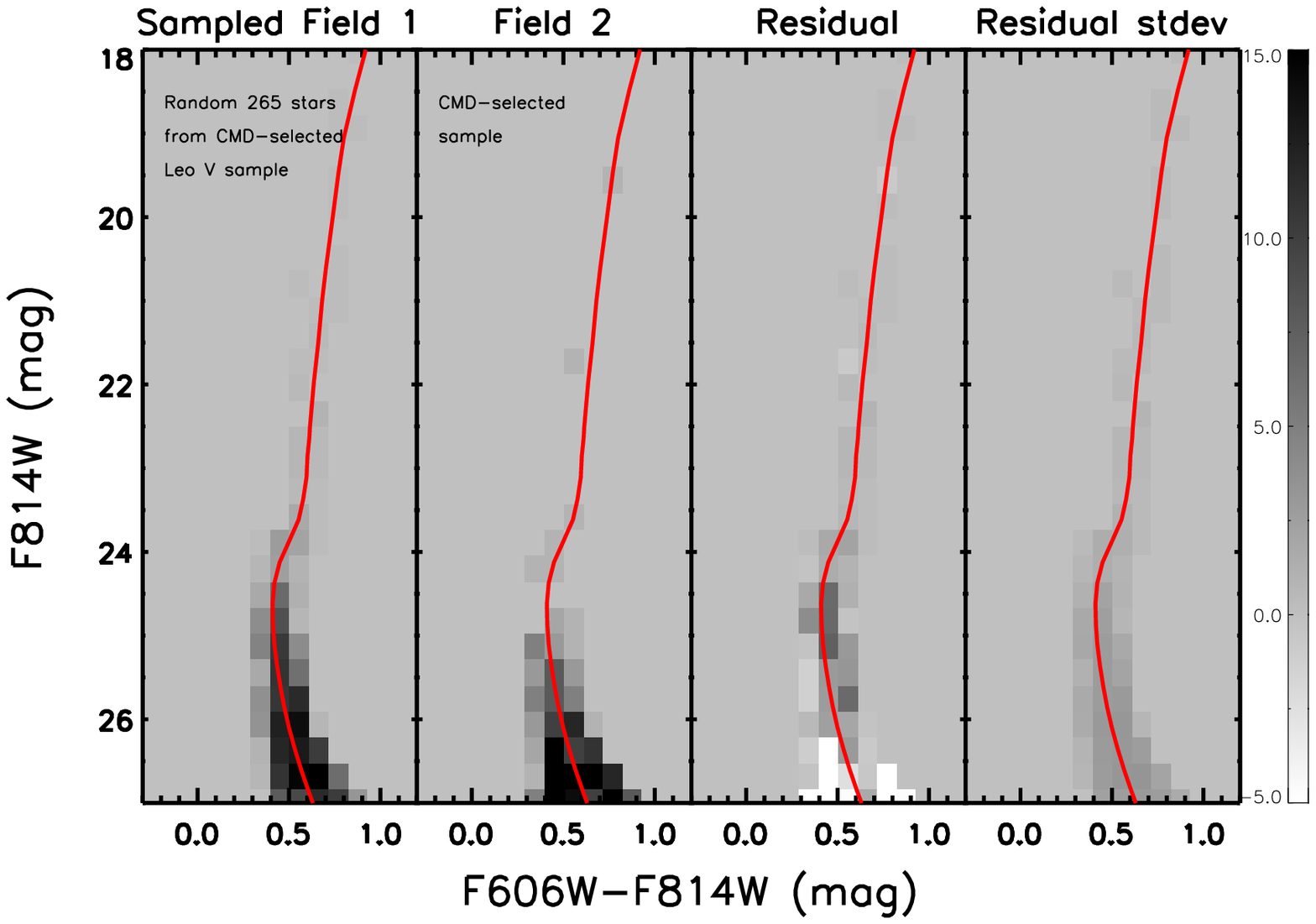}
    \caption{\textbf{Far-left:} Hess diagram of Field~1 ACS after randomly selecting 265 stars from its CMD-selected sample. \textbf{Center-left:} Hess diagram of the CMD-selected sample of Field~2 ACS. \textbf{Center-right:} Residuals after subtracting Field~2 (center-left panel) from sampled Field~1 (far-left panel). \textbf{Far-right:} Standard deviation of the residuals in our realizations. For reference, we show our M92 fiducial (red line). Field~2 seems to have an absence of bright stars and excess of faint main-sequence stars relative to Field~1.  \label{fig:nuggethess}}
\end{figure*}

\begin{figure*}[tbh!] 
\centering
\includegraphics[width=1.5\columnwidth]{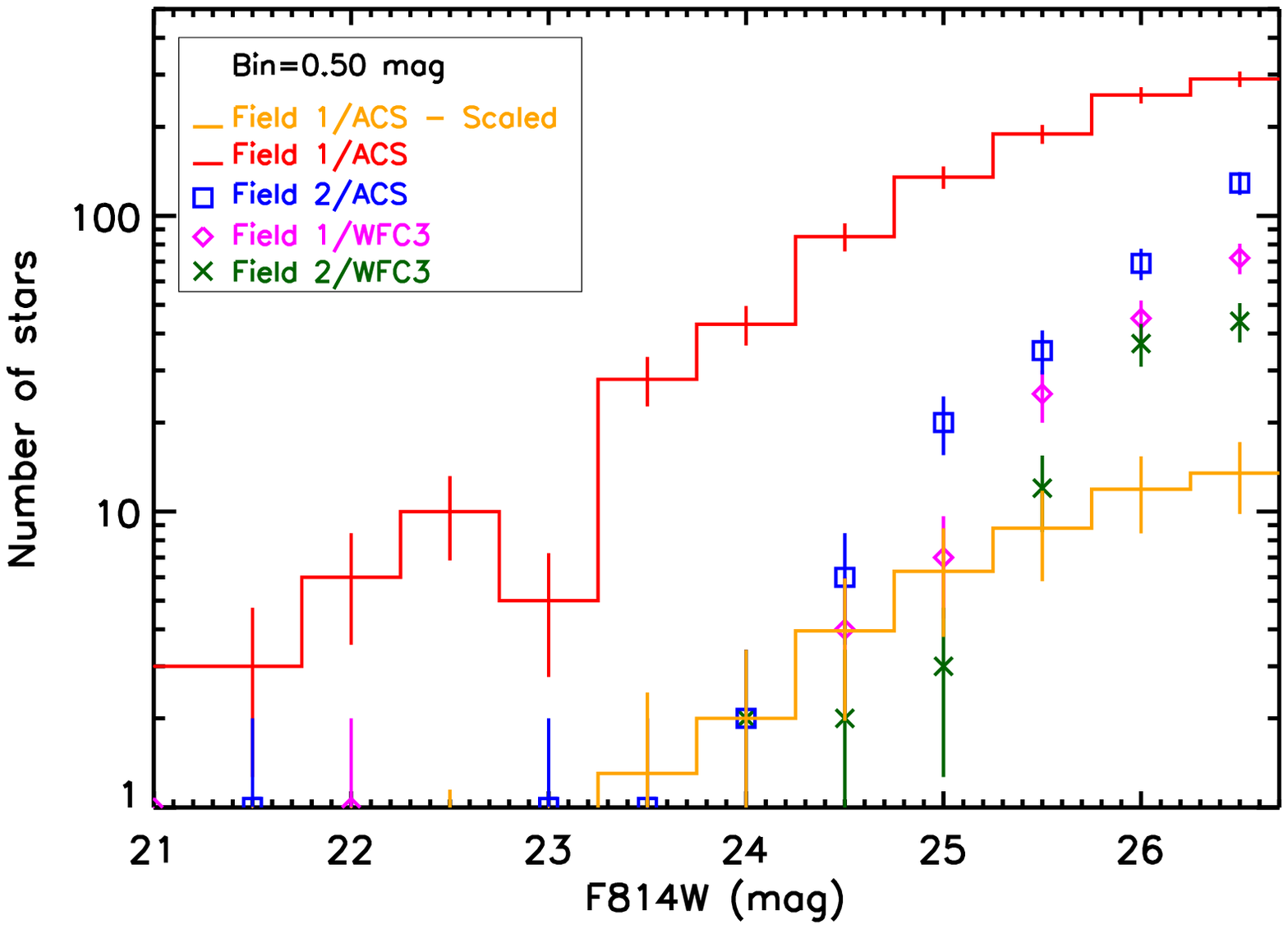}
\caption{Comparison of the observed luminosity functions of our \textit{HST} fields, utilizing only the CMD-selected stars. A scaled luminosity function of ACS Field~1 is plotted for a comparison, which has been normalized to match the number of ACS Field~2 stars at the F814W$=24.0$~mag bin. \label{fig:luminosityfunc}}
\end{figure*}

The stream-like overdensities of Leo~V were identified by \citet{Sand2012} in ground-based Megacam data, using a ``matched-filter'' algorithm \citep{Rockosi2002} that picks out stars consistent with an old, metal-poor stellar population in color-magnitude space. This Megacam data reaches point source depths (50\% completeness) of ($g$,$r$) $=$ ($26.20$, $25.75$)~mag. Here we investigate the \textit{HST}/ACS data of the candidate debris stream of Leo~V. Our main goal is to determine whether there is true tidal stellar debris associated with Leo~V. The \textit{HST}/WFC3 data is specifically used to provide control CMDs for positive or negative detections of tidal material, and to assess the impact of field contamination. We refer the reader to Figure~\ref{fig:obsplan} for our observational strategy. We also remind them that Field~1 is the central pointing of Leo~V and Field~2 is centered on the candidate debris stream. Finally, we note that our ACS fields have the same depth (see Table~\ref{tab:obslog}), hence they are comparable.


To investigate the apparent stellar overdensities, we focus on the region of Field~2 in the Megacam data, and select sources with colors and magnitudes expected for the stellar population of Leo~V. In our selection, we inflate the uncertainty to 0.1 mag (in order to account for the distance uncertainty) when the photometric errors are $<0.1$ mag. We then track down the CMD-selected Megacam objects in our \textit{HST} photometry. Almost half of those Megacam objects resolve out into background galaxies, highlighting the importance of \textit{HST} resolution for star/galaxy separation. 
The rest of the objects remain stars in the {\it HST} catalog, but many are no longer consistent with the stellar population of Leo~V with the improved accuracy of HST photometry. Thus, most candidate tidal stars in the Megacam data are actually background galaxies and nonmember foreground stars. 

We perform a CMD-selection using the M92 fiducial sequence on the Field~2 ACS data similar to the analysis we performed on the Megacam data. Within this selected population, there might be stars associated with Leo~V $-$either currently bound or in tidal material$-$ and field stars. The WFC3 fields are used here to assess the impact of field contamination, as they appear to have few Leo~V stars in the ground-based data, and they sample nearly the same Leo~V-centric radii as Field~2 (see Figure~\ref{fig:obsplan}). We test if the CMD of the identified sources in Field~2 agrees better with the CMD of Leo~V (Field~1) or those of the WFC3 fields. To do this, we apply our CMD-selection to the WFC3 fields, still  accounting for the photometric errors of each field. The red lines in Figure~\ref{fig:nuggetinspect} highlight the region selected by our M92 fiducial filter. While the center of Field~2 is $7.3\arcmin$ away from the center of Field~1 (Leo~V), the WFC3 fields are $6\arcmin$ and $9\arcmin$ for Field~1 and 2, respectively. When an ellipticity of $\sim$0.5 is assumed for Leo~V, these distances approximately translate into $12\arcmin$, $9.4\arcmin$, and $9.1\arcmin$ in elliptical radius ($r_{e}$), respectively. We note that the ellipticity of Leo~V is not well-constrained (its ellipticity is reported as $0.52\pm0.26$ by \citealt{Sand2012}, and $0.43\pm0.22$ by \citealt{Munoz2018}), and we consider the fields to be at similar projected radii for the purposes of this work.

Later, we create a Hess diagram of our ACS fields, utilizing only our CMD-selected stars in Figure~\ref{fig:nuggetinspect}, with a bin size of 0.3~mag in magnitude and 0.1~mag in color. The Field~2 sample has 265 stars with F814W$<27.0$~mag. To check the consistency to the stellar populations of Leo~V, we randomly select 265 stars from the Field~1 sample (1062 stars), and repeat this random selection 10,000 times. We note that performing the same analysis on stars with $24.0<$F814W$<27.0$~mag gives the same result. The far-left panel in Figure~\ref{fig:nuggethess} shows the average of these realizations in color-magnitude space. The center-left panel is the Hess diagram of the Field~2 sample (the original 265 stars mentioned above). The center- and far-right panels of this figure show the residuals after subtracting Field~2 (center-left panel) from the mean sampled Field~1 (far-left panel) and the standard deviation of the residuals in our realizations, respectively. In a true stellar debris, the sampled Leo~V CMD should be very similar to the Field~2 CMD, therefore one would expect the residuals in the center-right to have values of approximately zero or to be consistent with its standard deviations in the far-right. However, the lowest (highest) residual values reach $-$20 (8) stars per bin. The most significant residuals are associated with the mismatch of stars at $24.0\lesssim$F814W$\lesssim25.5$ and F814W$\gtrsim26.5$ between the stellar populations of Leo~V and Field~2, in which Field~2 seems to have an absence of bright stars and excess of faint main-sequence stars relative to Field~1. 
Note that the ACS fields have a very similar image depth (see Table~\ref{tab:obslog}), which makes our Hess diagrams comparable and the residual plots reliable. 

The mismatch between the CMD-selected samples of Field~1 and Field~2 is also visible in Figure~\ref{fig:luminosityfunc}, which compares the observed luminosity functions of our \textit{HST} fields (utilizing only the CMD-selected stars). We used a bin size of 0.5~mag, and errors bars are computed from the observed number of stars in each luminosity bin. For comparison, a scaled luminosity function of Field~1 ACS is plotted, which has been normalized to match the number of the Field~2 ACS stars at the F814W$=24.0$~mag bin. We remind the reader that the ACS and WFC3 fields have different image depths (see Table~\ref{tab:obslog}). While, at brighter magnitudes, the difference is almost negligible in the completeness and the photometric uncertainties, this consistency starts to break down at F814W $\gtrsim 26.5$~mag. Therefore, a comparison between the ACS and WFC3 fields can be trusted for magnitudes of F814W $< 26.5$~mag. The Field~2 stars show a better agreement to those of the WFC3 fields and an increasing mismatch with the Leo~V (Field~1) stars at fainter magnitudes. This casts doubt on the existence of a true tidal structure in Field~2, calling for attention in interpreting the overdensities in ground-based data. 
    
In short, our investigation on \textit{HST}/ACS and WFC3 observations of the candidate debris stream and surrounding regions shows that the CMD of the candidate debris stream is more consistent with our control field data, and not the stellar population of Leo~V itself, and we conclude that the overdensity detected in the Megacam data is not a true stellar stream, but instead the noise from unresolved background galaxies and stars with large photometric uncertainties. 

\section{Leo~V Members and Nonmembers}\label{sec:members}

In this section, we use stars with either spectroscopic measurements or {\it Gaia} proper motions, along with the known RR Lyrae stars \citep{Medina2017}, to identify Leo~V member or probable member stars.  This information will be used both to update the properties of Leo~V (e.g. proper motion), and to search for any signs of tidal disturbance in this data. 

\subsection{Previous Spectroscopic Studies}

There have been two previous spectroscopic studies of Leo~V.  The first accompanied the discovery of Leo~V, and used the MMT/Hectochelle spectrograph \citep{Walker2009}. We are presenting these spectra in the current paper using a uniform analysis with a complementary 2009 dataset, and will discuss them further below. Five likely members were identified within the central $\sim$3~arcmin of Leo~V (with a velocity dispersion of $\sigma=2.4^{+2.4}_{-1.4}$~km~s$^{-1}$), and two additional candidates were found at large radii ($R\sim$13~arcmin, or $\sim$700~pc). The second spectroscopic study was conducted by \citet{Collins2017} using DEIMOS on the 10-m Keck~II telescope, and found eight total member stars (with a velocity dispersion of $\sigma=2.3^{+3.2}_{-1.6}$~km~s$^{-1}$), three of which were in common with the initial \citeauthor{Walker2009} study, yielding five new members. One of them was later identified as a RR Lyrae (RRL) star along with two additional RRL stars \citep{Medina2017}. In addition, Collins and collaborators measured a tentative velocity gradient across the face of Leo~V in the direction of the Galactic center. This gradient, combined with the candidate spectroscopic member stars at large radii \citep{Walker2009}, and the photometric overdensities found in deep ground-based imaging \citep{Sand2012} all built the case that Leo~V may be disrupting.

\subsection{Combined Hectochelle Results}

We have collected Hectochelle data taken during two campaigns, one in 2008 and one in 2009, yielding 140 stellar observations that passed the quality-control criteria we outlined in Section~\ref{subsec:spectra}. These results are presented in Table~\ref{tab:hectoleo5}, including the derived $v$, $T_{eff}$, log $g$, and [Fe/H] from each observation.  

\begin{table*}[!hbtp]
\centering
\small
\caption{Previously identified `Leo~V' stars and comparison of their reported velocities.}\label{tab:velocity}
\begin{tabular}{cc|cc|cc|cc|c}
\tablewidth{0pt}
\hline
\hline
 \multirow{2}{*}{R.A. (deg)} & \multirow{2}{*}{Dec (deg)} & \multicolumn{2}{c}{This work} & \multicolumn{2}{|c}{\citeauthor{Walker2009}}  & \multicolumn{2}{|c|}{\citeauthor{Collins2017}} & \multirow{2}{*}{Comments} \\ 
  & &  {ID}           & {$v$ (km~s$^{-1}$)}   &   {ID}  & {$v$ (km~s$^{-1}$)} &  {ID}   & {$v$ (km~s$^{-1}$)} & \\
\hline
172.80503 &	2.21438 & \multicolumn{2}{c|}{not recovered}  & L5-1 &  $173.4\pm3.8$ & StarID-43 & $167.2\pm3.1$ & \nodata \\
172.79413 &	2.23597 &  LeoV-6   & $176.1\pm1.3$  &  L5-2 &  $174.8\pm0.9$ &  StarID-37 &  $173.3\pm2.3$ &  binary\\ 
          &         &           & $169.5\pm1.7$  &      &                &           &               &  \\     
172.76197 &	2.22080 &  LeoV-1   & $174.3\pm1.7$  & L5-4 & $173.2\pm1.5$  & StarID-25 & $177.8\pm2.3$ & \nodata\\
172.75690 & 2.19038 & \multicolumn{2}{c|}{not recovered} & L5-7 & $168.8\pm1.6$ & \nodata & \nodata & plausible  \\
172.81632 &	2.18272 &  LeoV-72  & $172.8\pm2.2$ & L5-8 & $176.8\pm2.1$  & \nodata & \nodata & \nodata\\
172.80879 &	2.44344 &  \multicolumn{2}{c|}{not recovered} &	L5-52 & $165.8\pm1.8$ & \nodata & \nodata & not reliable \\				
172.76729 &	2.44903 &  \multicolumn{2}{c|}{not recovered} &	L5-57 & $179.2\pm3.7$ & \nodata & \nodata & not reliable \\
172.73857 & 2.16262 &  \nodata  & \nodata       & \nodata & \nodata & StarID-17 & $173.0\pm3.7$ & \nodata\\				
172.77621 &	2.21669 &  \nodata  & \nodata       & \nodata & \nodata & StarID-27 & $170.8\pm3.2$ & \nodata\\				
172.77774 & 2.21724 &  \nodata & \nodata      & \nodata & \nodata & StarID-28 & $189.7\pm9.0$ &  RRL \\	
172.78563 & 2.21943 &  \nodata  & \nodata       & \nodata & \nodata & StarID-32 & $172.0\pm3.0$ & \nodata\\
172.80025 & 2.21661 &  \nodata  & \nodata       & \nodata & \nodata & StarID-41 & $164.4\pm2.5$ & \nodata\\
\hline
\end{tabular}
 \begin{tablenotes}
      \small
      \item NOTE $-$ Columns~1-2 are the right ascension, declination of the stars that were previously identified as `Leo~V' member in previous spectroscopic studies. We highlight a RRL star and a probable binary. See the text for the details. 
\end{tablenotes}
\end{table*}

\begin{table*}[!hbtp]
\centering
\footnotesize
\caption{Properties of Leo~V likely members}\label{tab:goodmembers}
\begin{tabular}{lcccccccc}
\tablewidth{0pt}
\hline
\hline
{No} & R.A. & Dec & $g$ & $r$ & $\mu_{\alpha} \cos{\delta}$ & $\mu_{\delta}$ & {Other IDs} & {Comment}\\
{} & (deg) & (deg) & (mag) & (mag) & (mas yr$^{-1}$) & (mas yr$^{-1}$)  &  {} & {}\\
\hline
1 &	172.80503 &	2.21438 &	21.02 &	20.38 &	$0.429\pm2.662$ & $-1.499\pm1.701$  & L5-1$^{a}$; StarID-43$^{b}$ & \nodata\\
2 &	172.79413 &	2.23597 &	20.61 &	19.83 &$-0.983\pm1.249$ & $-1.571\pm0.687$  & L5-2$^{a}$; StarID-37$^{b}$; LeoV-6$^{d}$ & binary\\
3 &	172.76197 &	2.22080 &	20.23 &	19.53 &$-0.91\pm1.041$ & $-0.189\pm0.555$   & L5-4$^{a}$; StarID-25$^{b}$; LeoV-1$^{d}$ & \nodata\\
4 &	172.81632 &	2.18272 &	20.26 &	19.52 &	$1.395\pm1.002$ &	$-0.586\pm0.559$ & L5-8$^{a}$; LeoV-72$^{b}$ & \nodata\\
5 &	172.73857 & 2.16262 & 21.55 &	20.97 & \nodata & \nodata  &  StarID-17$^{b}$ & \nodata\\				
6 &	172.77621 &	2.21669 & 22.09 &	21.48 & \nodata & \nodata  &  StarID-27$^{b}$ & \nodata\\				
7 &  172.77774 & 2.21724 & 21.98 & 21.71 & \nodata & \nodata  &   StarID-28$^{b}$; HiTS113107$+$021302$^{c}$ &  RRL \\	
8 & 172.78563 & 2.21943 & 21.08 &   20.49 & $1.837\pm3.173$ & $-3.838\pm2.203$ & StarID-32$^{b}$ & \nodata\\
9 & 172.80025 & 2.21661 & 20.52 &  19.83 & $-0.344\pm1.6$	& $-0.736\pm0.982$ & StarID-41$^{b}$ & \nodata\\
10 & 172.73946 & 2.22514 & 21.60 & 21.42 & \nodata & \nodata  &   HiTS113057$+$021330$^{c}$ & RRL \\
11 & 172.76936 & 2.22200 & 22.22 & 21.92 & \nodata & \nodata  &   HiTS113105$+$021319$^{c}$ & RRL \\
\hline
\end{tabular}
 \begin{tablenotes}
      \small
      \item $^a$ \citet{Walker2009}
      \item $^b$ \citet{Collins2017}
      \item $^c$ \citet{Medina2017}
      \item $^d$ This work
      \item NOTE $-$ Column 1 lists our assigned number for each star. Columns 2-5 are the right ascension, declination, $g$ and $r$-band magnitudes from \citet{Sand2012}, respectively. Columns 6-7 are the \textit{Gaia} DR2 proper motions. Column 8 lists other IDs for each star from the literature. We note Star~2 is a probable binary and Stars~7,10, and 11 are RR Lyrae stars. 
\end{tablenotes} 
\end{table*}

\begin{table*}[!hbtp]
\centering
\caption{List of plausible Leo~V members, requiring further follow-up to confirm.}\label{tab:maybemembers}
\begin{tabular}{lccccccc}
\tablewidth{0pt}
\hline
\hline
{No} & R.A. & Dec & $g$ & $r$ & $\mu_{\alpha} \cos{\delta}$ & $\mu_{\delta}$ & {Other IDs}\\
{} & (deg) & (deg) & (mag) & (mag) & (mas yr$^{-1}$) & (mas yr$^{-1}$)  &  {} \\
\hline
1p & 172.78508 & 2.3281389 & 19.72 & 19.11 & $-0.847\pm0.891$ & $-1.450\pm0.438$ & LeoV-99 \\
2p & 172.75690 & 2.1903850 & 20.42 & 19.67 & $-0.559\pm1.385$ & $-0.269\pm0.694$ & L5-7$^{a}$ \\
3p & 172.76526 & 2.34061 & 20.28 & 19.63 & $-2.803\pm1.244$ & $-0.832\pm0.633$ & \nodata\\
4p & 172.77572 & 2.25463 & 20.88 & 20.22 & $-2.053\pm2.783$ & $-2.743\pm1.507$ & \nodata \\
\hline
\end{tabular}
  \begin{tablenotes}
  \small
      \item[a] $^a$ \citet{Walker2009}
      \item NOTE $-$ Column~1 lists our assigned number for each star. Columns~2-5 are the right ascension, declination, $g$ and $r$-band magnitudes from \citet{Sand2012}, respectively. Columns~6-7 are the \textit{Gaia}~DR2 proper motions. We note that Star~1p is from our Hectochelle spectroscopic catalog while Stars~3p and 4p are stars without spectroscopy with consistent proper motions and consistent Megacam color-magnitudes.
\end{tablenotes}
\end{table*}

Here we first compare the \citeauthor{Walker2009} spectroscopy with the current work, focusing on likely Leo~V members. Of the five central stars considered to be Leo~V members in \citet{Walker2009}, we recover two of them with velocities consistent with the previous analysis to within the 2$\sigma$ uncertainties (see Table~\ref{tab:velocity}) -- these are labeled LeoV-1 and LeoV-72 in Table~\ref{tab:hectoleo5} (these are L5-4 and L5-8 in \citeauthor{Walker2009}, respectively). One of these, LeoV-1, was also observed by \citet{Collins2017} who also confirmed it to be a Leo~V member with a velocity about 1-$\sigma$ different from our own (StarID~25 in their Table~1). Two member stars in the original \citeauthor{Walker2009} analysis did not pass our quality cuts, and are not listed in Table~\ref{tab:hectoleo5} -- these are L5-1 and L5-7 using their parlance. One of these, L5-1, was also observed by \citet{Collins2017} who found it to be a Leo~V member star (StarID~43 from their work, at $V_{helio}$=167.2$\pm$3.1~km~s$^{-1}$). Both of these stars have proper motions consistent with the other Leo~V members, although at this point they make up a sizable fraction of Leo~V likely members and are possibly skewing the measured proper motion of the system; we discuss further in the next section. In the end, we believe that L5-1 (StarID~43) is likely a Leo~V member because of the \citeauthor{Collins2017} velocity measurement, but further spectroscopic measurements are warranted. However, L5-7 was not observed in the spectroscopic study of \citet{Collins2017}, and we no longer have confidence in its reported velocity. Given its imaging colors and consistent proper motion, we classify this star as a plausible candidate member of the system, an ideal target for future, deeper spectroscopic studies of Leo~V.

We confirm the fifth object identified by \citeauthor{Walker2009} (L5-2) to be a central Leo~V object, and we obtained a second epoch of spectroscopy in 2009. Our two epochs of spectroscopy of this Leo~V star, LeoV-6, yield significantly different results over that $\sim$1 year time span -- 176.1$\pm$1.3~km~s$^{-1}$ in 2008 and 169.5$\pm$1.7~km~s$^{-1}$ in 2009 -- and we consider this a likely binary star candidate. Observations by \citeauthor{Collins2017} yielded a $V_{helio}$=173.3$\pm$2.3~km~s$^{-1}$ for this object, confirming its association with Leo~V.

Two further Leo~V candidate members were identified by \citeauthor{Walker2009} -- L5-52 and L5-57 as presented in their Table~1 -- which were located $\sim$13~arcmin from the main body of Leo~V. These candidate members are of particular interest because they suggest a very extended stellar distribution for Leo~V, as they would be located $\gtrsim$10~half-light radii from its main body. In the new analysis of the 2008 dataset, however, neither object's spectra pass our quality control criteria. 
Neither object was observed in the spectroscopic study of \citet{Collins2017}, nor are they in the {\it Gaia} DR2 catalog. For these reasons, we no longer consider them Leo~V members, and this removes one of the observational clues that suggested Leo~V was tidally disrupting.

Given that we could not reproduce all of the results of \citet{Walker2009}, we also checked our velocity measurements using the cross-correlation method described in \citet{Johnson2017}, and find overall agreement between the different methods. We pay particular attention to those previously identified `Leo V' stars that did not pass our quality-control cuts. The cross-correlation technique gives a measured velocity of $\approx$173 km s$^{-1}$ for L5-7 and $\approx$175 km s$^{-1}$ for L5-57, in agreement with the previous work. In the case of L5-52, it only finds a consistent velocity ($\approx$168 km s$^{-1}$) for the 2009 data that \citeauthor{Walker2009} did not have. 
Meanwhile, we get consistent results for LeoV-6 via cross-correlation and the \citet{Walker2015} Bayesian analysis; hence, we are confident about each individual epoch for LeoV-6, which give reasonably strong evidence that this star is a binary. For L5-7, L5-52 and L5-57, definitive conclusions need further follow-up with better S/N, and we only present our trusted measurements using the \citet{Walker2015} technique. It goes without saying the measurement of high precision velocities for the Milky Way ultra-faint dwarfs is extremely challenging -- it may be in the future that multiple measurement techniques must be employed, and repeat measurements should be encouraged.  Many of the velocity measurements made of these systems over the last $\sim$15 years deserve a second look.

\begin{figure}[tbh!] 
\centering
\includegraphics[width=\columnwidth]{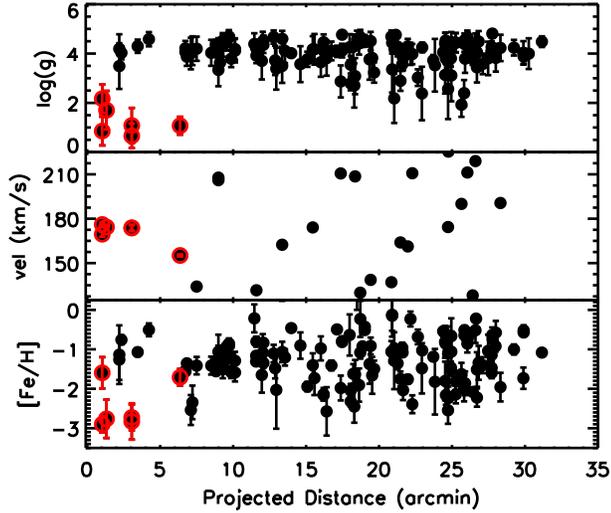}
\caption{Spectroscopically-measured surface gravity (top), velocity (middle) and metallicity (bottom) as a function of distance from Leo~V's center. Red markers represent stars that have velocity in the range of $150<v$ (km~s$^{-1}$) $<190$ (around the systemic velocity of $\sim$171~km~s$^{-1}$), low surface gravities (log~$g\lesssim2.5$) and low metallicities ([Fe/H]$\lesssim-1.5$), and they are well-separated from the foreground distribution in log~$g$ space. While the red data points within 5~arcmin are the individual observations of three kinematic members (LeoV-1, LeoV-6 and LeoV-72; with repeat measurements), the red marker at 6.4~arcmin is LeoV-99. \label{fig:specplot}}
\end{figure}

\begin{figure}[tbh!] 
\centering
\includegraphics[width=\columnwidth]{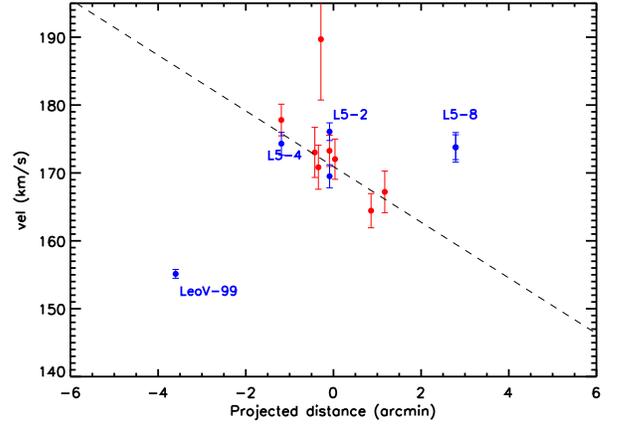}
\caption{Velocities as a function of projected distance along the kinematic major axis reported by \citet{Collins2017}. While blue filled circles are Leo~V members from \citet{Collins2017}, red ones are from our spectroscopic sample. LeoV-99 is shown as an open red circle. The dashed line represents the tentative kinematic gradient reported by \citet{Collins2017}. The position of the kinematic member LeoV-72 seems to weaken the velocity gradient argument, and our bootstrap analysis suggests that the reported kinematic gradients across Leo~V might be due to small number statistics.
\label{fig:velgrad}}
\end{figure}

Aside from the three kinematic member stars described above (LeoV-1, LeoV-6 and LeoV-72), five other member stars were identified by \citeauthor{Collins2017}, but none of these appear in our Hectochelle spectroscopic catalog. To search for new possible members in our Hectochelle sample, we plot surface gravity, velocity and metallicity values of each observation as a function of position (Figure~\ref{fig:specplot}). We highlight stars that have velocity in the range of $150<v$ (km~s$^{-1}$) $<190$ (around the systemic velocity of $\sim$171~km~s$^{-1}$), low surface gravities (log~$g\lesssim2.5$) and low metallicities ([Fe/H]$\lesssim-1.5$). In the log~$g$ panel, red points are well-separated from the foreground distribution, which has typically higher surface gravity values (e.g. foreground G dwarfs). Indeed, the measurements within 5~arcmin are the individual observations of three kinematic members (LeoV-1, LeoV-6 and LeoV-72; some with repeat measurements). The red marker at 6.4~arcmin is LeoV-99, which has a metallicity ([Fe/H]$=-1.7\pm 0.2$) consistent with Leo~V but has a velocity ($155.1\pm0.6$ km s$^{-1}$) offset from the systemic velocity ($170.9^{+2.1}_{-1.9}$~km~s$^{-1}$) found for Leo~V by \citet{Collins2017}. As it is located $\gtrapprox5$ half-light radii from the center of Leo~V, its lower velocity might be consistent with the reported velocity gradient. To investigate this further, we re-create Figure~6 of \citeauthor{Collins2017}, but this time also including kinematic members in our sample, and check the position of LeoV-99 (see Figure~\ref{fig:velgrad}). LeoV-99 does not fit into the picture expected from the velocity gradient. 
Further the position of kinematic member LeoV-72 seems to weaken the velocity gradient argument. We use a 10000 iteration bootstrap analysis to assess the significance of this gradient, and find a slope consistent with zero, suggesting that the reported kinematic gradients across Leo~V might be due to small number statistics. Interestingly, LeoV-99 also has a proper motion consistent with Leo~V but is blue-ward of the red giant branch, with $g\sim19$ and $g-r\sim0.6$~mag (see Figure~\ref{fig:gaia} and Table~\ref{tab:maybemembers}). As such, we do not classify LeoV-99 as a likely member, but highlight it as a plausible member of the system, requiring further follow-up to confirm. 

Given the refined spectroscopic membership presented in Table~\ref{tab:goodmembers}, it would be useful to re-derive the kinematic properties of Leo~V (e.g.,the systemic velocity, velocity dispersion). Unfortunately, after excluding the RR Lyrae velocity variables and our probable binary, we are left with seven stars and their velocity information comes from different data sets, in which there can be some systematic offset (see \citealt{Collins2017}). Of seven members, there is only one star in common, making it difficult to probe this possible velocity discrepancy. Therefore, we opt not to report any update on the kinematic properties based upon the available hybrid data. 

\subsection{{\it Gaia} Proper Motions}

\begin{figure*}[tbh!] 
\centering
\includegraphics[width=1.\columnwidth]{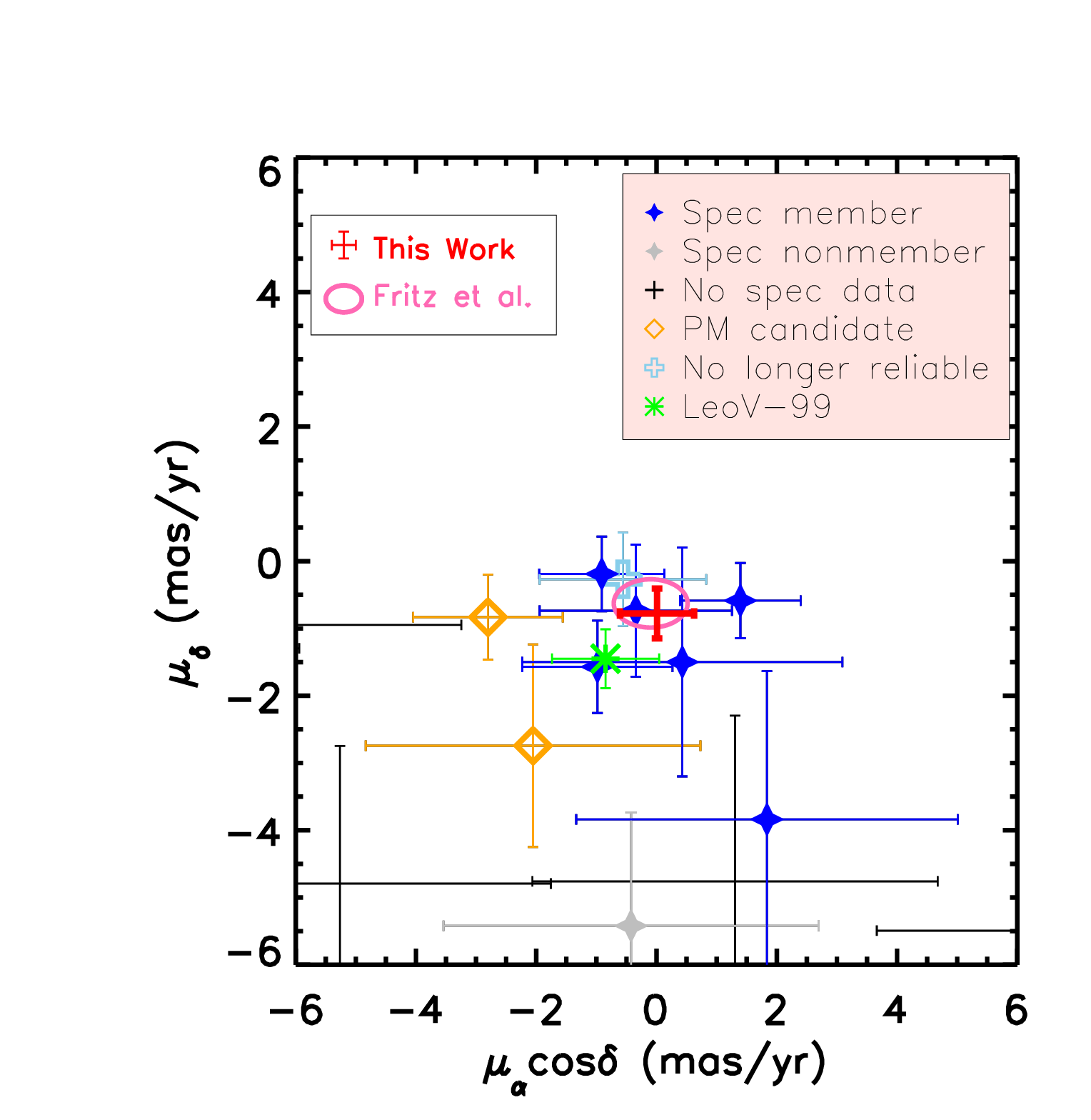}
\includegraphics[width=1.\columnwidth]{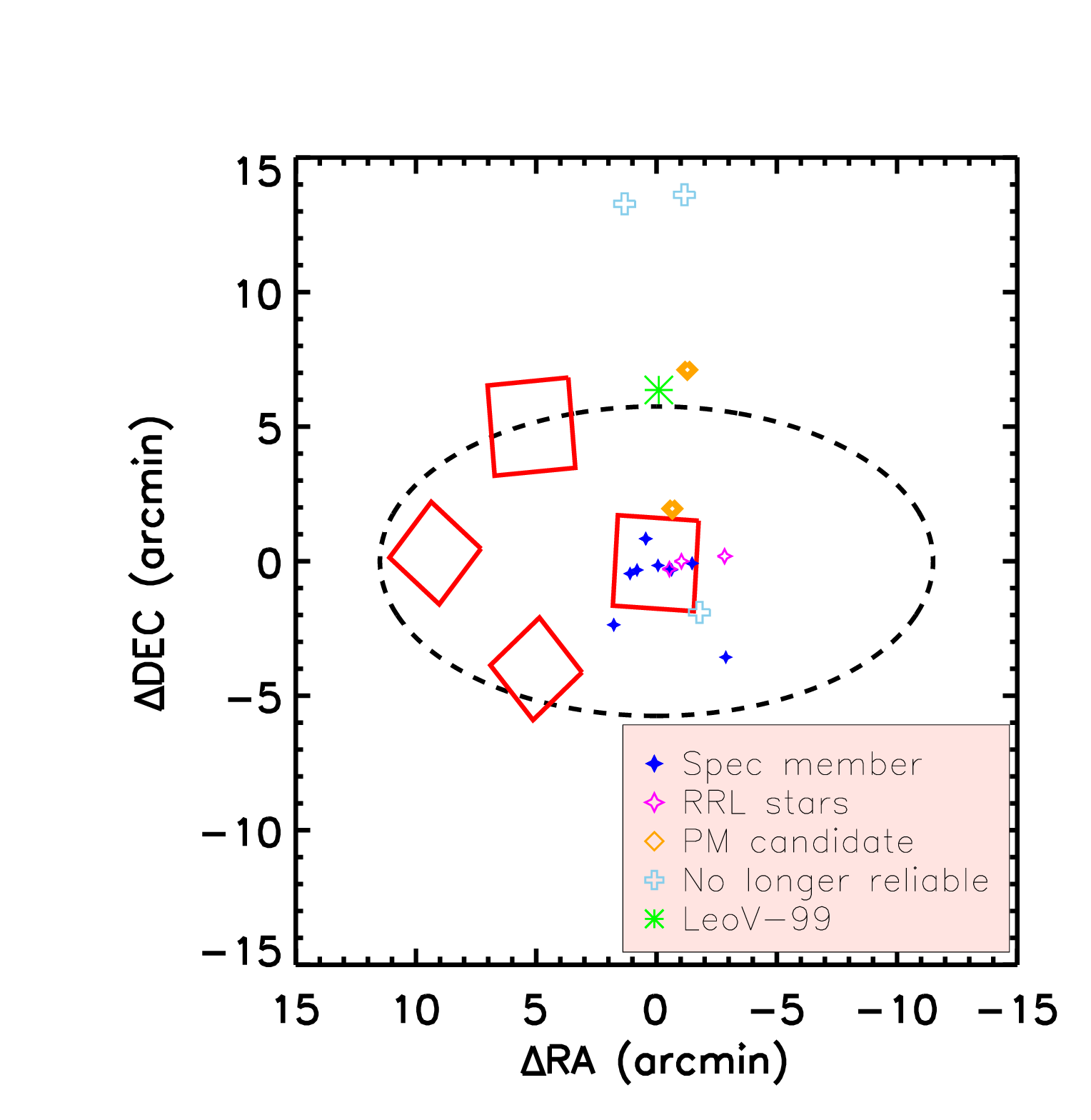}
\includegraphics[width=1.\columnwidth]{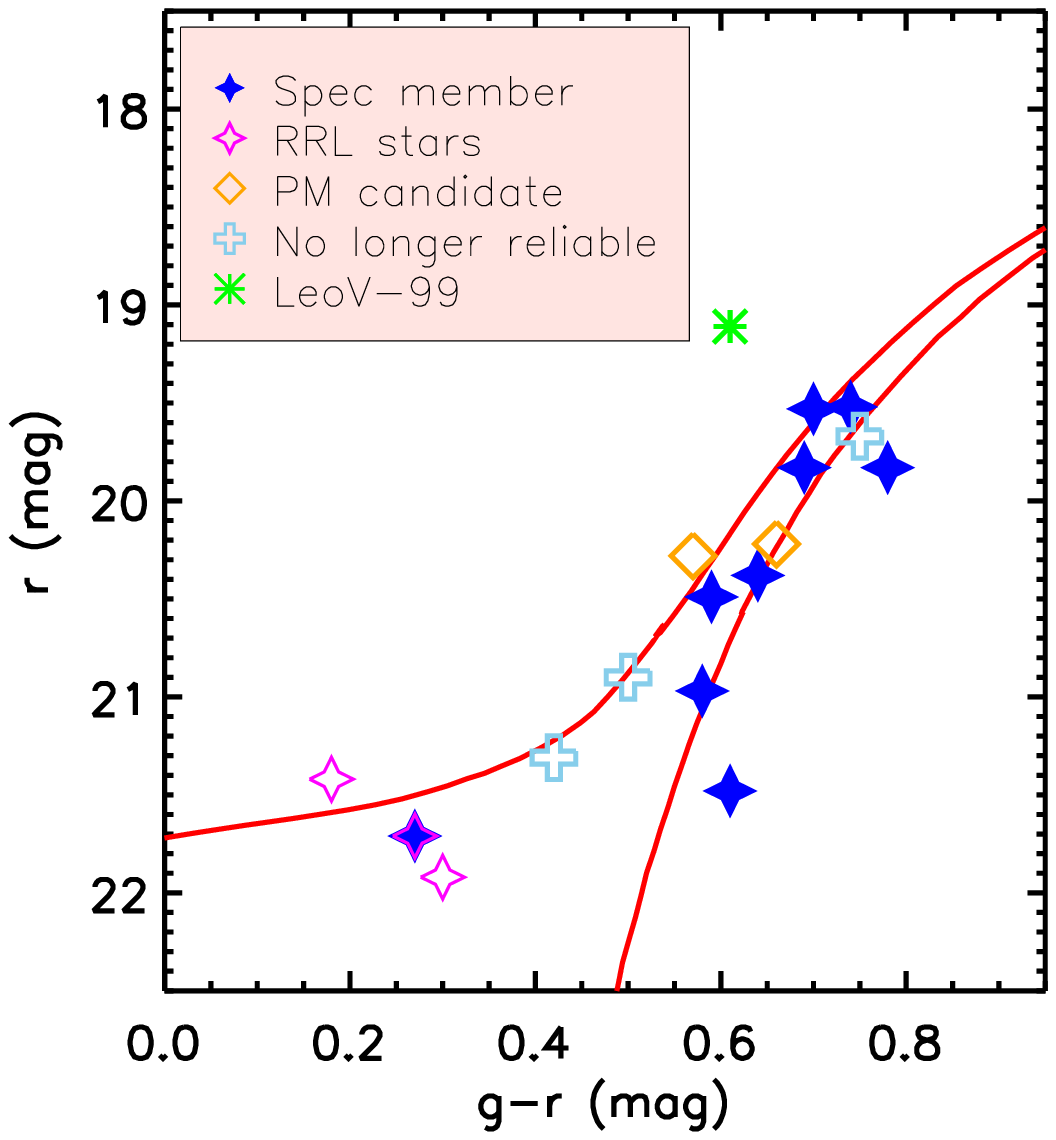}
\caption{\textbf{Top-left panel:} Proper motions of the CMD-selected stars in the Megacam data of \citet{Sand2012}. Within these stars, Leo~V spectroscopic likely members and nonmember stars are highlighted with blue and grey colors. Our proper motion estimation for Leo~V is shown in red, which is very close to the one reported by \citet{Fritz2018} (pink ellipse). Other symbols represent stars without spectroscopy, each defined with a particular legend. Light blue marker is L5-7, which has consistent PM values and colors but its velocity estimate is no longer reliable. \textbf{Top-right panel:} Locations of Leo~V likely members (spectroscopic members: blue, RRL: magenta) and plausible member of the system relative to the main body. Two distant light blue markers are L5-52 and L5-57, whose velocity estimates are no longer reliable.  \textbf{Bottom panel:} CMD of the stars in the top-right panel. Overplotted as a red line is the isochrone used in our CMD filter ([Fe/H]$=-2.0$, 13.5 Gyr, \citealt{Bressan2012}). \label{fig:gaia}}
\end{figure*}

\citet{Fritz2018} presented systemic proper motion (PM) measurements of dwarf galaxies which have been spectroscopically observed in the literature. Using the central five member candidates of Leo~V reported in \citeauthor{Walker2009}, the authors found $(\mu_\alpha \cos\delta, \mu_\delta)=(-0.097\pm0.557\pm0.057,-0.628\pm0.302\pm0.057)$, where the first uncertainty is the statistical error and the second one is the systematic error. The authors derived the systematic uncertainty by interpolating linearly between the values of 0.035 mas yr$^{-1}$ for sufficiently large objects on the sky \citep{Gaia2018} and 0.066 mas yr$^{-1}$ for smaller galaxies \citep{Lindegren2018}. Here, we use Leo~V likely members listed in Table~\ref{tab:goodmembers}, which includes three kinematic member stars from our Hectochelle spectroscopic catalog and member stars from \citeauthor{Collins2017}, and update the systemic PM measurement of Leo~V. Taking an error-weighted average of these six kinematic members, we find $(\mu_\alpha \cos\delta, \mu_\delta)= (0.009\pm0.560\pm0.057$,  $-0.777\pm0.314\pm0.057)$, where the first uncertainty is the statistical error and the second one is the same systematic error adopted by \citet{Fritz2018} for Leo~V. Our value is almost identical to the one reported by \citet{Fritz2018}, hence does not change the orbit of Leo~V, and it is still unlikely that Leo~V had a close encounter with the Milky Way.  

We search for further Leo~V member candidates using \textit{Gaia}~DR2, following a very similar methodology as that described in \citet{Carlin2018}. We use the Megacam photometry of Leo~V from \citet{Sand2012}, and select stars based on a CMD filter that isolates stars around an old, metal-poor isochrone ([Fe/H]$=-2.0$, 13.5 Gyr, \citealt{Bressan2012}) with width of 0.1 mag. Our selection includes all spectroscopic members. Then, we match these CMD-selected stars to \textit{Gaia}~DR2, and look for stars with PMs consistent with those of spectroscopic members. The top-left panel of Figure~\ref{fig:gaia} shows PMs of these CMD-selected stars. There are two new stars without spectroscopy with consistent PM values, which are labelled as possible PM candidates in the figure and listed as Stars~3p and 4p in Table~\ref{tab:maybemembers}. The top-right panel shows the locations of Leo~V likely members and plausible members of the system relative to the center of Leo~V. The red squares highlight our \textit{HST} pointings, and we also show the same black dashed ellipse in Figure~\ref{fig:obsplan}, as a reference. PM candidates also possess colors consistent with Leo~V membership (see the bottom panel). Interestingly, the PM candidate Star~3p is located close to LeoV-99. Assuming Leo~V is a uniformly sampled exponential sphere, the probability of finding two members at $\gtrapprox5$ half-light radii, in a sample of thirteen (Star~3p and LeoV-99 plus eleven likely members in Table~\ref{tab:goodmembers}), is $\sim2\times10^{-4}$. Thus, if LeoV-99 and Star~3p are real Leo~V members, this would imply tidal disruption. A spectroscopic study is required to really understand their association with Leo~V. 

\section{CONCLUSIONS}\label{sec:conclusion}

Previous studies suggest that Leo~V may be tidally disrupting. Ground-based observations \citep{Sand2012} show evidence of possible stripped stellar material far outside several half-light radii. The extended distribution of two member candidates of \citet{Walker2009} and the velocity gradients across Leo~V \citep{Collins2017}, together with its high ellipticity ($\sim0.5$), provided further evidence for tidal material. In this work, we present a combined {\it HST}, {\it Gaia}~DR2 and MMT/Hectochelle study on the tidal signatures seen in Leo~V. Here, we summarize our key results:

\begin{itemize}
    \item For a better understanding of the properties of Leo~V, we make a comparison with similar stellar systems $-$ Leo~IV and M92. We derive a new fiducial sequence for M92 (see Appendix~\ref{sec:AppendixA}), and revisit the dwarf's distance measurements, finding $m-M=21.25\pm0.08$ for Leo~V and $m-M=20.85\pm0.07$ for Leo~IV. Our comparison shows a close agreement between the CMDs of Leo~V, Leo~IV and M92, implying they have similar stellar populations and star formation histories. 

    \item Using \textit{HST}/ACS and \textit{HST}/WFC3 observations of the candidate debris stream and surrounding regions, we investigate whether there is true tidal stellar material associated with Leo~V. We find that the CMD of the candidate debris stream is more consistent with our control field data, and not the stellar population of Leo~V itself. Our work shows that caution is necessary for claims of tidal structures detected in ground-based data, and highlights the importance of deeper, high resolution observations for securing the detection of faint tidal features. 

    \item We present two epochs of Leo~V Hectochelle spectroscopy, one in 2008 and one in 2009. We revisit the first epoch of previously published 2008 data in order to analyze both in a uniform, robust fashion. The current results are largely consistent with that presented in \citet{Walker2009}. In the new analysis of the 2008 dataset, of five central stars considered to be Leo~V members, we recover two with velocities consistent with the previous analysis (LeoV-1 and Leo-72) and identify one as a probable binary with a velocity change of $\sim$7 km s$^{-1}$ ($\sim$4$\sigma$ detection) over $\sim$1 year (LeoV-6). Two other member candidates (L5-1 and L5-7) are no longer viable after our quality cuts, but both have proper motions consistent with the other Leo~V members. Given  L5-1's velocity measurement from \citet{Collins2017}, we still consider L5-1 as a likely member, but reclassify L5-7 as a plausible member of the system, requiring further follow-up to confirm.   

    
    \item In a search for member candidates in our spectroscopic catalog, we find one new plausible member, LeoV-99, which has a consistent surface gravity and metallicity, with a marginally consistent velocity as other Leo~V members. Its \textit{Gaia} proper motion value also implies Leo~V membership. Therefore, we consider this star worthy of being reported as a plausible candidate member of the system. In an effort to understand its membership status, we investigate the velocity gradient reported by \citet{Collins2017}. Inclusion of the kinematic members in our sample weakens the velocity gradient argument, casting doubt on another observational clue that suggested Leo~V was tidally disrupting.
    
    \item Six spectroscopic member candidates have \textit{Gaia} proper motions (see Table~\ref{tab:goodmembers}). Taking an error weighted average of these six stars, we find $(\mu_\alpha \cos\delta, \mu_\delta)= (0.009\pm0.560$,  $-0.777\pm0.314)$ mas yr$^{-1}$. This proper motion is very close to the one reported by \citet{Fritz2018}.
    
    \item In a search for Leo~V members using \textit{Gaia}~DR2, we find two new plausible candidates (without spectroscopy) with consistent colors and proper motions (see Table~\ref{tab:maybemembers}).  Interestingly, both candidates are located towards LeoV-99, and if confirmed would imply tidal disruption. They are ideal targets for a future spectroscopic study. 

\end{itemize}

Is Leo~V tidally disrupting? It is hard to settle the question. The kinematic membership of the two distant HB stars of \citet{Walker2009} is no longer reliable. Our {\it HST} investigation reveals that the candidate debris stream observed around Leo~V is likely not true tidal material. This calls into question the true nature of other stream-like overdensities around Leo~V. Also, we see evidence that the proposed kinematic gradient across Leo~V might be due to small number statistics. Overall, our findings dispute the case for disturbance in Leo~V. However, there are still things worth investigating: our new plausible distant members (LeoV-99 and Star~3p) may still support tidal disruption for Leo~V as these stars are located $>5$ half-light radii from the center of Leo~V.

The true nature of UFDs is hard to understand. As in our case, even a combined \textit{HST}, \textit{Gaia}~DR2 and MMT/Hectochelle study may not provide decisive results. We will continue our investigation of the signs of tidal disruption seen in several of the ultra-faint dwarf galaxies. The objects which have different observational features that suggest past Milky Way encounters (including extratidal stars, very high ellipticities, velocity gradients and strong deviations from the luminosity-metallicity relation) include Hercules, Ursa Major~I, Ursa Major~II, Segue~1, Segue~2, Tucana~III, Carina~I, Bo\"{o}tes~I, Willman~1, Crater~II,  and Draco~II -- the UFDs of the Milky Way are still far from being understood. A future \textit{HST}-based paper will be devoted to Hercules, which also shows evidence for extratidal stars \citep{Coleman07,Sand2009,MartinJin2010,Roderick2015,Garling2018} and a kinematic gradient \citep{Aden2009,Deason2012}. 

\acknowledgments

Support for this work was partly provided by NASA through grant number HST$-$GO$-15182.001$ from the Space Telescope Science Institute which is operated by AURA, Inc., under NASA contract NAS $5-26555$.
Research by DJS is supported by NSF grants AST-1821987, AST-1821967, AST-1813708, AST-1813466, and AST-1908972. MM is supported by NSF grants AST$-$1815403 and AST-1312997. EO is partially supported by NSF grant AST-1815767. JS acknowledges support from the Packard Foundation. We thank Ata Sarajedini for his help on accessing the ACS Globular Cluster Survey data. We also thank Andrew Dolphin for his help with DOLPHOT2.0. This work has made use of data from the European Space Agency (ESA) mission Gaia (\url{https://www.cosmos.esa.int/gaia}), processed by the Gaia Data Processing and Analysis Consortium (DPAC, \url{https://www.cosmos.esa.int/web/gaia/dpac/consortium}). 
Funding for the DPAC has been provided by national institutions, in particular the institutions participating in the Gaia Multilateral Agreement.  This research was supported in part by the National Science Foundation under Grant No. NSF PHY-1748958.  Part of this work was performed at the Aspen Center for Physics, which is supported by National Science Foundation grant PHY-1607611.

\vspace{5mm}
\facilities{HST (ACS, WFC3), Gaia, MMT (Hectochelle spectrograph)}

\software{
The IDL Astronomy User's Library \citep{IDLforever},
DOLPHOT2.0 \citep{Dolphin2002},
Topcat \citep{Taylor2005}
}

\bibliographystyle{aasjournal}
\bibliography{reference}

\appendix
\section{Fiducial Sequence of M92} \label{sec:AppendixA}

We present here a brief description of how we derive the fiducial sequence of M92. M92 was observed with \textit{HST}/ACS for two orbits on 2006 April 11 under the ACS Globular Cluster Treasury program (HST-GO-10775; PI: A. Sarajedini). This survey targeted the central regions of a large number of globular clusters, observing for one orbit in F606W and one orbit in F814W. The data-reduction procedure and the resulting catalog for each cluster were described in \citet{Anderson2008}. We use their M92 photometry with the updated photometric zeropoints \citep{Mack2007}\footnote{The M92 {\it HST}/ACS photometry is taken from \url{https://archive.stsci.edu/prepds/acsggct/}}. We implement the extinction corrections for M92 using the same method described in Section~\ref{sec:obs}, with an average $E(B-V)$ of $0.022$ mag. 

\begin{figure}[htbp!] 
\centering
\includegraphics[width=\columnwidth]{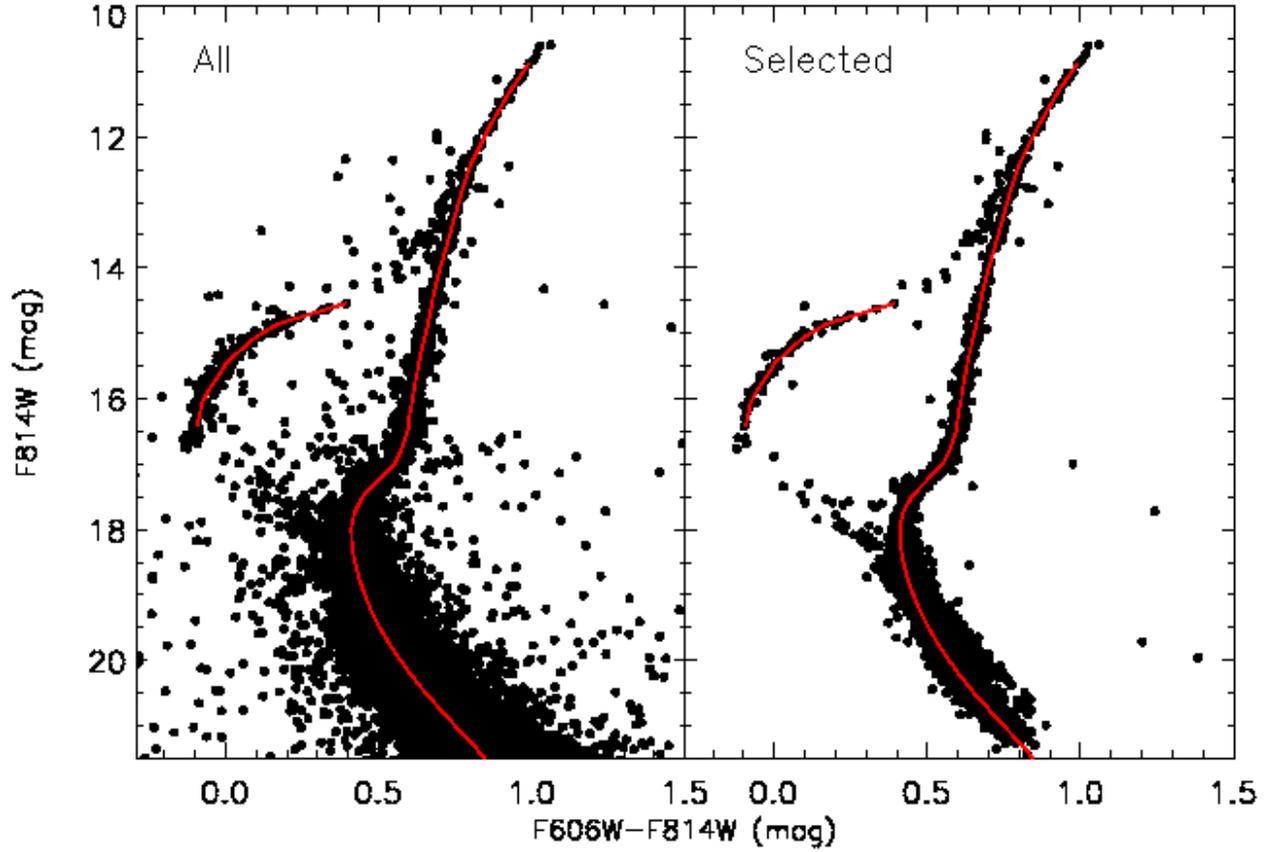}
\caption{Left: CMD of all the stars in the M92 catalog of \citet{Anderson2008}. Right: CMD of those stars judged to have the highest quality photometry on the basis of their $qfit$, $oth$, $xsig$ and $ysig$ values as described in the text. It is important to mention that the right panel does not represent a complete sample of M92, but it shows a representative sample for the derivation of the fiducial sequence. Overplotted as a red line is our fiducial sequence of M92.
\label{fig:m92cmd}}
\end{figure}

The M92 catalog includes some general measurement-quality information, some of which are the quality of the PSF-fit in each filter ($qfit$: smaller is better), fraction of light in the aperture due to neighbors ($oth$), and the difference between the x and y positions in different filters ($xsig$, $ysig$). The column-by-column description for the catalog can be found in Table~4 of \citet{Anderson2008}. To produce a well-defined cluster sequence, we use these measurement-quality metrics to judge which stars have the highest quality photometry. Specifically, we include only those stars with $qfit$~$\leq0.05$ and $oth$~$\leq0.1$ in each filter, $xsig$~$\leq 0.005$ and $ysig$~$ \leq 0.005$ for absolute magnitude $M_{F814W} \leq -1.0$ mag (assuming a distance modulus of $m-M=14.62$ mag for M92). Figure~\ref{fig:m92cmd} shows two CMDs for M92 with all the stars in the catalog in the left panel and those stars that survived our imposed cuts in the right. We note that the majority of stars that are excluded in the right panel  are due to the effects of crowding, which would otherwise result in a noisier main sequence and turnoff region. Our selected sample does not represent a complete sample of M92, but it provides a well-defined and very tight cluster sequence extending from the red-giant branch to approximately four magnitudes below the turnoff point. 

We follow a very similar methodology as that described in \citet{Clem2008}: an empirical ridge line is derived by determining the median color of stars that lie within different F814W magnitude bins, adopting larger magnitude bins along the main-sequence and red-giant branch, and smaller bins for the subgiant branch. Our fiducial sequence is overplotted as a red line in Figure~\ref{fig:m92cmd} and tabulated in Table~\ref{tab:fiducial}. Note that the M92 photometry used in this paper and our fiducial sequence are in the VEGAMAG system \citep{Sirianni2005} while the photometry of \citet{Brown2014} and their fiducials \citep{Brown2005} are in the STMAG system. 

\begin{table*}[ht!]
\centering
\caption{Extinction Corrected Ridge Lines for the Globular Cluster M92 in the VEGAMAG system} \label{tab:fiducial}
\begin{tabular}{cc}
\tablewidth{0pt}
\hline
\hline
F606W    & F814W \\
(mag)    & (mag) \\
\hline
11.88 &	10.89\\
12.23 &	11.30\\
12.71 &	11.85\\
13.22 &	12.42\\
13.60 &	12.83\\
14.12 &	13.38\\
14.67 &	13.97\\
15.05 &	14.37\\
15.53 &	14.87\\
15.98 &	15.34\\
16.49 &	15.88\\
16.62 &	16.01\\
16.83 &	16.22\\
17.07 &	16.48\\
17.32 &	16.74\\
17.54 &	16.98\\
17.74 &	17.24\\
17.94 &	17.49\\
18.16 &	17.74\\
18.40 &	17.99\\
18.65 &	18.24\\
18.90 &	18.48\\
19.17 &	18.73\\
19.43 &	18.98\\
19.69 & 19.22\\
19.97 &	19.47\\
20.26 &	19.73\\
20.54 &	19.98\\
20.83 & 20.22\\
21.12 &	20.48\\
21.56 &	20.84\\
22.16 &	21.34\\
22.71 &	21.80\\
23.12 &	22.16\\
\hline
\end{tabular}
\end{table*}


\end{document}